\documentclass[a4paper,12pt]{iopart}

\usepackage{iopams,graphicx,cite}

\newenvironment{eqnArray}{\begin{equation}\begin{array}[b]{rcl}}
{\end{array}\end{equation}}  %% implements the usual eqnarray environment
\newcommand{\ds}{\displaystyle}
\newcommand{\INT}[1]{%
  \int\limits_{\mbox{\small$\scriptstyle({#1}\geq0)$}}\!\!(\D{#1})}
\newcommand{\Eq}[1]{(\ref{eq:#1})}
\newcommand{\I}{\mathalpha{\mathrm{i}}}
\newcommand{\D}{\mathalpha{\mathrm{d}}}
\newcommand{\half}{\frac{1}{2}}
\newcommand{\pdi}[2]{\frac{\partial#2}{\partial#1}}
\newcommand{\expectn}[1]{\langle#1\rangle}
\newcommand{\Expect}[2][0pt]{\mathbb{E}{\left(#2\rule{0pt}{#1}\right)}}
\renewcommand{\tr}[2][0pt]{\mathrm{tr}{\left(#2\rule{0pt}{#1}\right)}}
\newcommand{\determ}[1]{\mathrm{det}{\left(#1\right)}}
\newcommand{\peak}[1][\rho]{#1_{\mathrm{peak}}}
\newcommand{\ML}{\rho^{\ }_{\mathrm{ML}}}
\newcommand{\LINV}{\widehat{\rho}}
\newcommand{\adj}{^{\dagger}}\newcommand{\phadj}{^{\vphantom{\dagger}}}
\newcommand{\bcol}[2][c]{\bigl(\begin{array}{#1}#2\end{array}\bigr)}
\newcommand{\column}[2][c]{{\left(\begin{array}{#1}#2\end{array}\right)}}

\newcommand{\power}[1]{^{\mbox{\footnotesize$#1$}}}
\newcommand{\Exp}[1]{\mathalpha{\mathrm{e}}^{\mbox{\footnotesize$#1$}}}
\newcommand{\unit}{\mathbf{1}}
\newcommand{\normal}{\mathalpha{\mathrm{N}}}
\newcommand{\ccW}{\mathalpha{\mathrm{CCW}}}
\newcommand{\quW}{\mathalpha{\mathrm{W}}^{(\mathrm{Q})}}
\newcommand{\Rand}[1]{\underline{#1}}
\newcommand{\vts}{\rule{0.1em}{0ex}}  %% very thin space
\newcommand{\nqb}{n_{\mathrm{qb}}}

\newcommand{\Lp}[1][1.1]{\mathopen{\scalebox{#1}{$($}}}
\newcommand{\Rp}[1][1.1]{\mathclose{\scalebox{#1}{$)$}}}
\newcommand{\Chi}{\mathalpha{\raisebox{0.4ex}{\scalebox{1.2}{$\chi$}}}}
\newcommand{\BL}{\mathopen{\scalebox{1.2}{$($}}}
\newcommand{\BR}{\mathclose{\scalebox{1.2}{$)$}}}
\newcommand{\ExpecT}[2][1.2]{\mathbb{E}%
  \mathopen{\scalebox{#1}{$($}}{#2}\mathclose{\scalebox{#1}{$)$}}}
\newcommand{\estc}{\widehat{\rule{0.07em}{0ex}c_{\lambda}\rule{0.06em}{0ex}}}
\newcommand{\ests}{\widehat{\rule{0.07em}{0ex}s_{\lambda}\rule{0.06em}{0ex}}}
\newcommand{\ssum}{\mathop{\scalebox{0.9}{$\ds\sum$}}}
\newcommand{\sint}{\mathop{\scalebox{0.9}{$\ds\int$}}}

\DeclareMathAlphabet{\vecfont}{OT1}{cmr}{bx}{it}
\renewcommand{\vec}[1]{\vecfont{#1}}
\renewcommand{\submitto}[1]{\vspace{28pt plus 10pt minus 18pt}
  \noindent{\small\rm Posted on the arXiv on #1.}}

\begin{document}

\title{Uncorrelated problem-specific samples of quantum states %
       from zero-mean Wishart distributions}

\markboth{R Han \textit{et al.}: %
       Uncorrelated samples of quantum states (zero-mean Wishart)}{}

\author{Rui Han$^{1}$, %
  Weijun Li$^1$\footnote{Now at %
    Trinity Hall, Cambridge University, Cambridge CB3 0D2, UK $\not\in$ EU}, %
    Shrobona Bagchi$^1$\footnote{Now at %
      Raymond and Beverly Sackler School of Physics and Astronomy, %
      Tel-Aviv University, Tel-Aviv 69978, Israel}, %
    Hui Khoon Ng$^{2,1,3}$ and Berthold-Georg Englert$^{1,3,4}$}

\address{$^1$Centre for Quantum Technologies, %
             National University of Singapore, Singapore}
\address{$^2$Yale-NUS College, Singapore}
\address{$^3$MajuLab, CNRS-UCA-SU-NUS-NTU International Joint Research Unit, %
             Singapore} 
\address{$^4$Department of Physics, National University of Singapore, %
             Singapore}
\ead{han.rui@quantumlah.org}

\begin{abstract}
Random samples of quantum states are an important resource for various tasks
in quantum information science, and samples in accordance with a
problem-specific distribution can be indispensable ingredients.
Some algorithms generate random samples by a lottery that follows certain
rules and yield samples from the set of distributions that the lottery can
access.
Other algorithms, which use random walks in the state space, can be tailored
to any distribution, at the price of autocorrelations in the sample and with
restrictions to low-dimensional systems in practical implementations.
In this paper, we present a two-step algorithm for sampling from the quantum
state space that overcomes some of these limitations.

We first produce a CPU-cheap large proposal sample, of uncorrelated
entries, by drawing from the family of complex Wishart distributions, and
then reject or accept the entries in the proposal sample such that the
accepted sample is strictly in accordance with the target distribution.  
We establish the explicit form of the induced Wishart distribution for quantum
states.
This enables us to generate a proposal sample that mimics the target
distribution and, therefore, the efficiency of the algorithm, measured by the
acceptance rate, can be many orders of magnitude larger than that for a
uniform sample as the proposal.

We demonstrate that this sampling algorithm is very efficient for one-qubit
and two-qubit states, and reasonably efficient for three-qubit states, while
it suffers from the ``curse of dimensionality'' when sampling from structured
distributions of four-qubit states. 
\end{abstract}


\submitto{16 June 2021}

\maketitle %% equivalent to \newpage; has no other effect

%%% switch to numbered footnotes in the main text
\makeatletter%
\setcounter{footnote}{0}%
\renewcommand{\thefootnote}{\arabic{footnote}}%
\def\@makefnmark{\hbox{$^{\thefootnote}\m@th$}}%
\long\def\@makefntext#1{\parindent 1em\noindent 
  \footnotesize\rm$\m@th^{\thefootnote}$~#1}%
\makeatother%
%%% 

\section{Introduction}\label{sec:intro}
Random states --- states with random properties --- play a central role in
statistical mechanics because one simply cannot account for the huge number of
dynamical degrees of freedom in systems composed of very many constituents.
In quantum mechanics --- in particular, in quantum information applications ---
random states also appear naturally even if there are relatively few degrees
of freedom.
On the one hand, there is always the noisy and uncontrollable interaction of
the quantum system under study with its environment;
on the other hand, our knowledge about the system is always uncertain because
the state parameters cannot be known with perfect precision.

There are many instances in quantum information science where a random sample
of quantum states, pure or mixed, is desirable:
in the evaluation of entanglement
\cite{Horodecki2008,Hamma2012,Dupuis2015,Collins2016};
for the numerical testing of a noise model for state preparation or gate
operation \cite{Sim2020,Lu2020};
when optimizing a function on the state space \cite{Collins2016};
for determining the size and credibility of a region in the state space
\cite{Li2016,Oh2019a,Oh2019b}; 
when determining the minimum output entropy of a quantum channel
\cite{Collins2016}; and in many other applications.
While direct sampling, in which an easy-to-generate raw sample is
converted into a sample from the desired distribution by an appropriate
transformation, is an option for some particular distributions, it does not
offer a flexible general strategy.

The seemingly simple requirement that the quantum state is represented by a
positive unit-trace Hilbert-space operator translates into complicated
conditions obeyed by the variables that parameterize the state, and these
constraints must be enforced in the sampling algorithm.
In one standard approach, the probability distribution of the quantum states
is defined by the sampling algorithm --- a lottery for pure system-and-ancilla
states, for example, followed by tracing out the ancilla
\cite{Zyczkowski2011,Collins2016,Bengtsson2017} --- and this can yield large
high-quality samples for the distributions that are accessible in this way.
This includes uniform distributions for the Haar measure, the 
Hilbert--Schmidt measure, the Bures measure, or
the measure induced by the partial trace
\cite{Braunstein1996,Hall1998,Bengtsson2017}.%
\footnote{The sample used in \cite{Gu2019} for demonstrating that local
  hidden variables are a lost cause is of this lottery-defined kind, too.}
The major drawback is that the exact probability distribution for random
states constructed this way are typically not known and one has to rely on the
error-prone numerical estimation for an approximated distribution.

Another standard approach generates samples by a random walk in the quantum
state space (or, rather, in the parameter space for it).
These Monte Carlo (MC) algorithms can accommodate any
\emph{target distribution} by flexible rules for the walk.
The well-known Markov Chain MC algorithm can generate samples in the
probability simplex, followed by an accept-reject step that enforces
the constraints \cite{Liu2008, Shang2015}.
There is also the Hamiltonian MC algorithm, in which the constraints are
obeyed by design, at the price of a rather more complicated parameterization
\cite{Neal2011, Seah2015}.%
\footnote{The samples of quantum channels in \cite{Sim2020} are generated in
  this way.}
These random-walk algorithms yield samples with autocorrelations, and the
effective sample size can be a small fraction.
Moreover, we encounter implementation problems when applying the MC algorithm
to higher-dimensional system.

In view of these observations, two directions suggest themselves for
exploration. 
One could begin with easy-to-generate and flexible raw samples
drawn from a \emph{proposal distribution}, richer
in structure than the uniform samples mentioned above, and then perform
suitable acceptance-rejection resampling (or \emph{rejection sampling} in
standard terminology) to obtain a sample in accordance with
the target distribution.
Or one could process large uncorrelated%
\footnote{The term ``uncorrelated'' is synonymous to ``independent and
    identically distributed'' (i.i.d.), which is standard terminology in
    statistics.}
raw samples by tailored random walks that eventually enforce the constraints.
In this work, we explore the first suggestion by adapting the complex Wishart
distributions%
\footnote{So named in recognition of John Wishart's
  \cite{Pearson1957} seminal work of 1928 \cite{Wishart1928}.
  For important follow-up developments and extensive discussions of Wishart
  distributions as well as, more generally, of random matrices and positive
  random matrices as tools for multivariate statistical analysis with
  real-life applications in, for example, biology, physics, social science,
  and meteorology,
  see \cite{Goodman63,Srivastava65,Muirhead82,Johnson02,Anderson03}.} 
from the statistics literature to suitable proposal distributions of quantum
states. 
The second suggestion is the subject matter of a separate paper where we use a
sequentially constraint MC algorithm~\cite{Li2020}.

The general idea of the first suggestion --- generate a sample drawn from the
target distribution by rejecting or accepting entries from a larger
CPU-cheap raw sample --- has many potential 
implementations, and the target distribution will inform the choice we make
for the proposal distribution from which we draw the raw sample.
We demonstrate the practical feasibility of the method at the example of
target distributions as they arise in the context of quantum state estimation
(section~\ref{sec:target}), and we use raw samples generated from properly
adapted Wishart distributions (section~\ref{sec:proposals}).
The target and proposal distributions used for illustration have single narrow
peaks, but that is no limitation as convex sums of several Wishart
distributions are proposal distributions for target distributions with more
peaks.

More specifically, we introduce the complex Wishart distribution of a quantum
state in section~\ref{sec:proposals} and derive its explicit probability
distribution, followed by a look at its important properties, such as the peak
location and peak shape; then we discuss how to get a useful proposal
distribution from it.
Section~\ref{sec:target} deals with the exemplary target distributions that
one meets in quantum state estimation.
In section~\ref{sec:accept-reject}, then, we show how rejection
sampling converts a sample drawn from the proposal distribution into a sample
drawn from the target distribution.
We verify the sample so generated by the procedure described in
section~\ref{sec:verify}, with technical details delegated to the appendix. 
After commenting on the limitations that follow from the ``curse of
dimensionality'' in section~\ref{sec:curse}, we present various samples of
quantum states in section~\ref{sec:examples} that illustrate our sampling
method. 
The examples demonstrate that the method is reliable and competitive but does
not escape the curse of dimensionality.
We close with a summary and outlook in section~\ref{sec:summary}.

\section{Wishart distributions of quantum states}\label{sec:proposals}
\subsection{The quantum state space}\label{sec:q-state-space}
{A state $\rho$ of a $m$-dimensional quantum system, represented by a
positive-semidefinite unit-trace $m\times m$ matrix, is of the form
\begin{equation}
  \label{eq:rho-varrho}
  \rho=\frac{1}{m}\unit_m+\sum_{l=1}^{m^2-1}\varrho_l B_l\,,
\end{equation}
where $\unit_m$ is the $m\times m$ unit matrix, the set
$\bigl\{B_1,B_2,\dots,B_{m^2-1}\bigr\}$ is an orthonormal basis in the
$(m^2-1)$-dimensional real vector space of traceless hermitian $m\times m$
matrices,
\begin{equation}
  \label{eq:Bbasis}
  B_l\phadj=B_l\adj\,,\quad\tr{B_l}=0\,,\quad\tr{B_{l}B_{l'}}=\delta_{ll'}\,,
\end{equation}
and the $\varrho_l$s are the coordinates for the vector associated with
$\rho$.
The volume element in this euclidean space,
\begin{equation}\label{eq:drho}
  (\D\rho)=\prod_{l=1}^{m^2-1}\D\varrho_l\,,
\end{equation}
is invariant under basis changes;
it is independent of the choice made for the $B_l$s.
Since $\tr{\rule{0pt}{10pt}(\rho-\rho')^2}=\sum_l(\varrho_l-\varrho_l')^2$,
this is the volume element induced by the Hilbert--Schmidt distance.
If we use the first $m-1$ diagonal matrix elements $\rho_{jj}$ of $\rho$ as
well as the real and imaginary parts of the off-diagonal matrix elements,
${\rho_{jk}^{\ }=\rho_{jk}^{(\mathrm{r})}+\I\rho_{jk}^{(\I)}=\rho_{kj}^*}$
for $j<k$, as integration variables, which are coordinates in a basis that is
not orthonormal, we have 
\begin{eqnarray}\label{eq:drho'}
  (\D\rho)={\left(2^{m(m-1)}m\right)}^{\frac{1}{2}}[\D\rho]
  \quad\mbox{with}\quad
  [\D\rho]=\prod_{j=1}^{m-1}\D\rho_{jj}\prod_{k=j+1}^m
              \D\rho^{(\mathrm{r})}_{jk}\,\D\rho^{(\I)}_{jk}\,,
\end{eqnarray}
where the prefactor is the Jacobian determinant.
The euclidean space contains quantum states  (${\rho\geq0}$) and also
unphysical $\rho\vts$s that have negative eigenvalues (${\rho\not\geq0}$).
The total volume of the convex quantum state space is stated in
\Eq{qu-state-volume}. 

In preparation of the following discussion of sampling $\rho$ from a Wishart
distribution, we note that the $m\times m$ matrix for a quantum state can be
written as 
\begin{equation}
  \label{eq:rho-Psi}
  \rho=\frac{\Psi\Psi\adj}{\tr{\Psi\Psi\adj}}
  \quad\textrm{with}\quad\Psi=\bcol[cccc]{\psi_1&\psi_2&\cdots&\psi_{n}}\,,
\end{equation}
where $\Psi$ is a $m\times n$ matrix composed of $n$ $m$-component columns.
If we view the columns of $\Psi$ as representations of pure states, then
\begin{equation}
  \label{eq:rho-psis}
  \rho=\frac{\psi_1\phadj\psi_1\adj+\psi_2\phadj\psi_2\adj
    +\cdots+\psi_{n}\phadj\psi_{n}\adj}
  {\psi_1\adj\psi_1\phadj+\psi_2\adj\psi_2\phadj
    +\cdots+\psi_{n}\adj\psi_{n}\phadj}
\end{equation}
is a mixed state blended from $n$ pure states, \emph{as if} a pure
system-and-ancilla state were marginalized by tracing out the ancilla.
The rank of $\rho$ cannot exceed $\min\{m,n\}$ and is usually equal to this
minimum when $\Psi$ is chosen ``at random,'' that is: drawn from a
distribution on the state space, as discussed in the subsequent sections.
For the purposes of this paper, we take ${n\geq m}$ for granted.

\subsection{The Wishart distributions}\label{sec:Wishart}
\subsubsection{Gaussian distribution of complex matrices.}
Now, to give meaning to the phrase ``$\,\Psi$ is chosen at random,''
we regard the real and imaginary parts of the matrix elements
${\Psi_{jk}=\alpha_{jk}+\I\beta_{jk}}$ of $\Psi$ as coordinates in a
$2mn$-dimensional euclidean space with the volume element
\begin{equation}
  \label{eq:dPsi}
  (\D\Psi)=\prod_{j=1}^m\prod_{k=1}^{n}\D\alpha_{jk}\,\D\beta_{jk}\,,
\end{equation}
and then have a multivariate, zero-mean, gaussian distribution on this matrix
space specified by the probability element
\begin{equation}
  \label{eq:G}
  (\D\Psi)\, G(\Psi|\Sigma)=(\D\Psi)\,(2\pi)^{-mn}\,\determ{\Sigma}^{-n}\,
  \Exp{-\frac{1}{2}\tr{\Psi\adj\Sigma^{-1}\Psi}}\,,
\end{equation}
where $\Sigma$ is a positive $m\times m$ matrix --- the covariance matrix for
each column in $\Psi$.
Denoting the random variable by $\Rand{\Psi}$ and one of its values by $\Psi$,
we say that ``$\,\Rand{\Psi}$ is drawn from the zero-mean normal distribution
of $m\times n$ matrices with the $m\times m$ covariance matrix $\Sigma\,$''
and, adopting the conventional notation from the statistics literature (see
\cite{Wasserman2004}, for example), write
${\Rand{\Psi}\sim\normal_{mn}(0,\unit_{n}\otimes\Sigma)}$.%
\footnote{More general gaussian distributions have
  $\tr{\Sigma_2^{-1}(\Psi-\Phi)\adj\Sigma_1^{-1}(\Psi-\Phi)}$ in the exponent,
  with two covariance matrices $\Sigma_1$ and $\Sigma_2$ and a displacement
  matrix $\Phi$, and then
  ${\Rand{\Psi}\sim\normal_{mn}(\Phi,\Sigma_2\otimes\Sigma_1)}$. 
  We only need the basic gaussian of \Eq{G} here, and explore the option of
  $\Phi\neq0$ in a companion paper \cite{Bagchi2020}.}
Since, in analogy with \Eq{rho-psis}, we can read the $n\times n$ trace as
\begin{equation}
  \label{eq:trnxn}
  \tr{\Psi\adj\Sigma^{-1}\Psi}=\sum_{k=1}^{n}\psi_k\adj\Sigma^{-1}\psi_k\phadj\,,
\end{equation}
the exponential function in \Eq{G} is the product of $n$ exponential
functions, one factor for each column of $\Psi$.
Having noted this factorization, we can state in which sense $\Sigma$ is  the
covariance matrix: For each $m$-component random column $\psi$ 
we have the expected values ${\Expect{\psi}=0}$ for the mean and
${\Expect{\left(\rule{0pt}{10pt}\psi-\Expect{\psi}\right)%
    \left(\rule{0pt}{10pt}\psi-\Expect{\psi}\right)\adj}%
    =\Expect{\psi\psi\adj}=\Sigma}$ for the covariance.
It follows that ${\Expect{\Psi\Psi\adj}=n\Sigma}$.

\subsubsection{Wishart  distribution of hermitian  matrices.}
The numerator in \Eq{rho-Psi} or \Eq{rho-psis} is a hermitian $m\times m$
matrix ${R=\Psi\Psi\adj}$ with matrix elements
${R_{jk}=x_{jk}+\I y_{jk}=R_{kj}^*}$
with which we associate the $m^2$-dimensional euclidean space with the volume
element
\begin{equation}\label{eq:dR}
(\D R)=\prod_{j=1}^m\D x_{jj}\prod_{k=j+1}^m\D x_{jk}\,\D y_{jk}\,.  
\end{equation}
As stated in Theorem 5.1 in \cite{Goodman63}, the induced probability
distribution on this $R$ space has the probability element
\begin{eqnarray}
  \label{eq:W}
  (\D R)\,W(R|\Sigma)
  &=&(\D R)\int(\D\Psi)\,G(\Psi|\Sigma)\,\delta(R-\Psi\Psi\adj)
                         \nonumber\\
  &=&(\D R)\,\frac{\determ{R}^{n-m}}
                  {2^{mn}\,\Gamma_m(n)\,\determ{\Sigma}^{n}}
      \,\Exp{-\frac{1}{2}\tr{\Sigma^{-1}R}}
\end{eqnarray}
where
\begin{equation}
  \label{eq:tildeGamma}
  \Gamma_m(n)=\prod_{j=0}^{m-1}\pi^j \Gamma(n-j)
             =\pi^{\frac{1}{2}m(m-1)}\prod_{j=1}^{m}(n-j)!
\end{equation}
is a multivariate gamma function.
The random variable $\Rand{R}$ is drawn from the centered complex Wishart
(CCW) distribution of hermitian $m\times m$ matrices
\cite{Wishart1928,Goodman63,Srivastava65}, 
${\Rand{R}\sim \ccW_m(n,\Sigma)}$, that derives from
${\normal_{mn}(0,\unit_{n}\otimes\Sigma)}$.%
\footnote{When ${n<m}$, all $R\vts$s are rank deficient and the
  anti-Wishart distribution is obtained instead \cite{Yu2002,Janik2003}.
  We do not consider samples of rank-deficient matrices here.}

\subsubsection{The quantum Wishart distribution.}
% {Wishart  distribution of quantum states.}
The conversion of this distribution for $R$ to a distribution for
${\rho=R/\tr{R}}$ requires an integration over ${\tau=\tr{R}}$ after expressing
the $R_{jk}$s in terms of $\tau$ and the $\rho_{jk}$s, omitting $\rho_{mm}$
which does not appear in \Eq{drho'}, that is
\begin{eqnArray}\label{eq:R-rho}
  R_{jk}&=&\rho_{jk}\tau \quad\mbox{for}\quad (j,k)\neq(m,m)\,,\\[1ex]
  R_{mm}&=&\displaystyle{\left(1-\sum_{j=1}^{m-1}\rho_{jj}\right)}\tau\,,
\end{eqnArray}
so that the volume elements in $R$ space and $\rho$ space are related by
\begin{equation}\label{eq:dR-drho}
  (\D R)=[\D\rho]\,\D\tau\,\tau^{m^2-1}\,.
\end{equation}
Accordingly, upon evaluating the $\tau$ integral in
\begin{equation}
  \label{eq:g'}
  (\D\rho)\,g(\rho)
  =[\D\rho]\,\int_0^{\infty}\D\tau\,\tau^{m^2-1}W(R=\tau\rho|\Sigma)
\end{equation}
we arrive at the following \textbf{Theorem:}%
\footnote{The case of $\Sigma=\unit_m$, when
  ${g(\rho)\propto\determ{\rho}^{n-m}}$ accounts in full for the $\rho$
  dependence, is well known; see, for example, (15.58) in \cite{Bengtsson2017}
  or (3.5) in \cite{Zyczkowski2001}.}
\begin{equation}\label{eq:g}
  \parbox{0.64\textwidth}{%\centerline{\bfseries Theorem}%
    For ${\Rand{\Psi}\sim\normal_{mn}(0,\unit_{n}\otimes\Sigma)}$
    with ${n\geq m}$, the $m\times m$ matrices
    \mbox{\small$\ds{\rho=\frac{\Psi\Psi\adj}{\tr{\Psi\Psi\adj}}}$}
    are drawn from the Wishart distribution of $m$-dimensional quantum states,
    ${\Rand{\rho}\sim\quW_m(n,\Sigma)}$, which is specified by the probability
    element
    \begin{center}$\ds%
      (\D\rho)\,g(\rho)=[\D\rho]\,
      \frac{\Gamma(mn)}{\Gamma_m(n)}
      \frac{\determ{\rho}^{n-m}}{\determ{\Sigma}^{n}\,
        \tr{\Sigma^{-1}\rho}^{mn}}\,.
    $\end{center}}
\end{equation}
By construction, ${g(\rho)=0}$ for all unphysical $\rho\vts$s.
For the sake of notational simplicity, we leave the dependence of $g(\rho)$ on
$m$, $n$, and $\Sigma$ implicit, just as we do not indicate the dependence of
$G(\Psi|\Sigma)$ on $m$ and $n$ in~\Eq{G}.

We observe that, when drawing a $\Psi$ from the distribution $G(\Psi|\Sigma)$
to arrive at a quantum state $\rho$ drawn from the distribution $g(\rho)$, it
does not matter if we replace  $\Sigma$ by $\lambda\Sigma$ with a positive
scaling factor $\lambda$ because the product
$\determ{\lambda\Sigma}\,\tr[10pt]{(\lambda\Sigma)^{-1}\rho}^{m}$ does not
depend on~$\lambda$.
We write ${\Sigma'\doteq\Sigma}$ for two $\Sigma$s that are
equivalent in this sense.

When ${\Sigma\doteq\unit_m}$, the probability distribution $g(\rho)$
is isotropic in the sense of ${g(U\rho U\adj)=g(\rho)}$ for all unitary
$m\times m$ matrices $U$. 
In the case of a generic ${\Sigma>0}$, the distribution is
invariant under ${\rho\to U\rho U\adj}$ if $U$ commutes with $\Sigma$,
and only then. 

\subsubsection{Uniformly distributed quantum states.}
A particular case is that of ${n=m}$ and ${\Sigma\doteq\unit_m}$, when
$g(\rho)$ is constant and equal to the reciprocal of the volume occupied
by the quantum states in the $(m^2-1)$-dimensional euclidean space,
\begin{equation}\label{eq:qu-state-volume}
  \INT{\rho}={\left(2^{m(m-1)}m\right)}^{\half}
  \frac{\Gamma_m(m)}{\Gamma(m^2)}
  %=\frac{(2\pi)^{\half m(m-1)}m^{\half}}{(m^2-1)!}\prod_{j=0}^{m-1}j!
  \,.
\end{equation}
Here, the $\rho\vts$s are uniformly distributed over the state space ---
uniform in the sense of the Hilbert--Schmidt distance.%
\footnote{It appears that, for many authors, uniformity in the
  Hilbert--Schmidt distance is the default meaning of ``picking a quantum
  state at random;'' an example is the marginalization performed in
  \cite{Faist2016}.} 
This provides a computationally cheap and reliable way of generating
an uncorrelated, uniform sample of quantum states of this kind.

\subsection{Efficient sampling}\label{sec:effSample}
In addition to unitary transformations of $\rho$, we can also consider general
hermitian conjugations, which invites the following question:
If $\rho$ is drawn from the distribution $g(\rho)$, what is the
corresponding distribution for
\begin{equation}\label{eq:rho->rho'}
  \rho'=\frac{A\rho A\adj}{\tr{A\rho A\adj}}\,,
\end{equation}
where $A$ is a nonsingular $m\times m$ matrix?
Clearly, the sample of $\rho'$s can be generated by the replacement
${\Psi\to\Psi'=A\Psi}$ for each $\Psi$ drawn from $G(\Psi|\Sigma)$.
We write $A$ in its polar form,
\begin{equation}\label{eq:A-polar}
  A=HU\,,
\end{equation}
where $U$ is a unitary matrix and ${H=(AA\adj)^{\frac{1}{2}}}$ is a positive
matrix, and note that
\begin{equation}\label{eq:dPsi'}
  (\D\Psi')=(\D\Psi)\,\determ{H}^{2n}
\end{equation}
for the respective volume elements in \Eq{dPsi}.
Then,
\begin{eqnarray}\label{eq:G'}
  (\D\Psi)\,G(\Psi|\Sigma)
  &=&(\D\Psi')\,\determ{H}^{-2n}G(A^{-1}\Psi'|\Sigma)\nonumber\\
  &=&(\D\Psi')\,G(\Psi'|A\Sigma A\adj)\,,
\end{eqnarray}
so that $\Psi'$ is drawn from the multivariate gaussian distribution with the
covariance matrix $A\Sigma A\adj$.%
\footnote{This is the statement of Theorem 1.2.6 in \cite{Muirhead82}.}
Here, then, is the answer to the question above:
$\rho'$ is drawn from $g(\rho')$ with $\Sigma$ replaced by
$A\Sigma A\adj$ in 
\Eq{g}, that is ${\Rand{\rho'}\sim\quW_m(n,A\Sigma A\adj)}$.
Note that ${\quW_m(n,A\Sigma A\adj)=\quW_m(n,\Sigma)}$ when
${A\Sigma A\adj\doteq\Sigma}$, that is:
${A=\lambda^{\half}\Sigma^{\half}U\Sigma^{-\half}}$ with $\lambda$ a positive
number and $U$ a unitary matrix. 
This includes the case, discussed above, of a unitary $A$ that commutes with
$\Sigma$. 

This observation is of practical importance.
We can sample $\Psi$ from $G(\Psi|\unit_m)$ --- drawing both the real and the
imaginary parts of all matrix elements $\Psi_{jk}$ from the one-dimensional
gaussian distribution with zero mean and unit variance, which can be done very
efficiently ---
and then put ${\rho=A\Psi\Psi\adj A\adj/\tr{A\adj A\Psi\Psi\adj}}$ with $A$
such that ${\Sigma\doteq AA\adj}$; the choice ${A=\Sigma^{\frac{1}{2}}}$
suggests itself. 
This yields a random sample of uncorrelated $\rho\vts$s drawn from $g(\rho)$.

\subsection{Peak location}\label{sec:peak}
In \Eq{g}, we have
\begin{equation}\label{eq:g''}
  g(\rho)\propto\frac{\determ{\rho_{\Sigma}^{\ }}^{n}}{\determ{\rho}^m}
  \quad\mbox{with}\quad
      \rho_{\Sigma}^{\ }=\frac{\Sigma^{-\frac{1}{2}}\rho\Sigma^{-\frac{1}{2}}}
                         {\tr{\Sigma^{-1}\rho}}\,,
\end{equation}
so that $g(\rho)$ is peaked at the $\rho$ that compromises between maximizing
$\determ{\rho_{\Sigma}^{\ }}$ and minimizing $\determ{\rho}$.
Since the response of $\determ{\rho}$ to an infinitesimal change of $\rho$ is
\begin{equation}
  \label{eq:delta-det}
  \delta\,\determ{\rho}=\determ{\rho}\,\tr{\rho^{-1}\delta\rho}\,,
\end{equation}
the stationarity requirement $g(\peak+\delta\rho)=g(\peak)$ reads
\begin{equation}
  \label{eq:delta-g}
  \tr{{\left[(n-m)\peak^{-1}
        -\frac{mn\Sigma^{-1}}{\tr{\Sigma^{-1}\peak}}\right]}\delta\rho}=0
\end{equation}
for all permissible $\delta\rho\vts$s.
It follows that we need%
\footnote{For a hermitian $m\times m$ matrix $A_{\rho}$ that is a functional
  of $\rho$, the requirement that ${\tr{A_{\rho}\delta\rho}=0}$ holds for all
  permissible   $\delta\rho\vts$s implies
  ${A_{\rho}\rho=\rho A_{\rho}=\rho\,\tr{A_{\rho}\rho}}$.}
\begin{equation}
  \label{eq:optSigma}
  \Sigma\doteq{\left(\peak^{-1}+\frac{m^2}{n-m}\unit_m\right)}^{-1}
\end{equation}
if we want $g(\rho)$ to be largest at $\rho=\peak$,
\begin{equation}
  \label{eq:g-max}
  \max_{\rho}{\left\{g(\rho)\right\}}=g(\peak)\,.
\end{equation}
Clearly, for every full-rank $\peak$, there are probability densities $g(\rho)$
that have their maximum there, one such $g(\rho)$ for each $n$ value larger
than $m$.

The covariance matrix \Eq{optSigma} applies for ${n>m}$ and a full-rank
$\peak$, as ${g(\rho)=0}$ for a rank-deficient $\rho$ when ${n>m}$.
If ${n=m}$, we get ${\Sigma\doteq\unit_m}$ for all full-rank $\peak$s, which
is the ${g(\rho)=\mathrm{constant}}$ situation noted above.

For ${n=m}$ and ${\Sigma\not\doteq\unit_m}$, $g(\rho)$ is largest when
$\tr{\Sigma^{-1}\rho}$ is smallest and, therefore, the eigenvalue-subspace of
$\Sigma$ for its largest eigenvalue comprises all $\peak$s.
While this enables us to design $g(\rho)$ such that it is maximal for a
rank-deficient $\peak$, it is of little practical use; more about this in
section~\ref{sec:target2}.

\subsection{Peak shape}\label{sec:shape}
For $\rho\vts$s near $\peak$,
\begin{equation}
  \label{eq:rho-eps}
  \rho=\peak+\epsilon=\peak+\sum_{l=1}^{m^2-1}\varepsilon_lB_l\,,
\end{equation}
where the traceless hermitian $m\times m$ matrix $\epsilon$
and its real coordinates $\varepsilon_l$ are small on the relevant scales,
we have 
\begin{eqnarray}
  \label{eq:peak-shape}
  \ds\log\frac{g(\peak+\epsilon)}{g(\peak)}
  &\cong&\ds
  -\half(n-m){\left[\tr{{\left(\peak^{-1}\epsilon\right)}^2}
      -\frac{n-m}{mn}\tr{\peak^{-1}\epsilon}^2\right]}\nonumber\\
  &=&\ds-\half\sum_{l,l'}\varepsilon_l G_{ll'} \varepsilon_{l'}
\end{eqnarray}
with
\begin{eqnarray}
  \label{eq:shape-G}
  G_{ll'}&=&(n-m)\tr{\peak^{-1}B_l\peak^{-1}B_{l'}}\nonumber\\
  &&\mbox{}-\frac{(n-m)^2}{mn}\tr{\peak^{-1}B_l}\tr{\peak^{-1}B_{l'}}
\end{eqnarray}
upon discarding terms proportional to the third and higher powers of
$\epsilon$. 
For the gaussian approximation of $g(\rho)$ in the vicinity of $\peak$ in
\Eq{peak-shape}, the inverse of the matrix $G$ is the covariance matrix in the
coordinate space. 
It follows that the distribution $g(\rho)$ is narrower when $n$ is larger;
and since $\peak^{-\half}\epsilon\peak^{-\half}$ appears in \Eq{peak-shape},
rather than $\epsilon$ by itself, the distribution will be particularly narrow
in the directions associated with the smallest eigenvalues of $\peak$. 

We break for a quick look at one-qubit and multi-qubit
examples and then pick up the story in section~\ref{sec:linshift}.

\subsection{Example: Wishart samples for a qubit}\label{sec:oneQB}
The qubit case (${m=2}$) is the one case that we can visualize.
The state space is the unit Bloch ball \cite{Bloch1946} with cartesian
coordinates $x,y,z$ introduced in the standard way,
\begin{eqnArray}\label{eq:QB-1}\fl
  \rho&=&\ds\frac{1}{2}\column[cc]{1+z & x-\I y \\ x+\I y & 1-z}
  =\half\bigl(\unit_2+x\sigma_x+y\sigma_y+z\sigma_z\bigr)
  =\half(\unit_2+\expectn{\bsigma}\cdot\bsigma)\,,   \\[3ex]{}
  \hspace*{-2em}
  (\D\rho)&=&\ds\frac{1}{\sqrt{8}}\,\D x\,\D y\,\D z\,,\quad
  \determ{\rho}=\frac{1}{4}(1-x^2-y^2-z^2)\,,
\end{eqnArray}
where $\sigma_x$, $\sigma_y$, $\sigma_z$ are the standard Pauli matrices, the
cartesian components of the vector matrix $\bsigma$, and $x$, $y$, $z$ are
their respective expectation values,
such as $x=\expectn{\sigma_x}=\tr{\rho\sigma_x}$,
with ${x^2+y^2+z^2\leq1}$.
Any choice of right-handed coordinate axes is fine (proper rotations in the
three-dimensional euclidean space are unitary transformations of the
$2\times2$ matrices) and, 
by convention, we choose the coordinate axes such that $\Sigma$ is diagonal,
\begin{equation}
  \label{eq:QB-2}
  \Sigma\doteq\column[cc]{\Exp{\vartheta} & 0 \\ 0 & \Exp{-\vartheta}}\,,
  \quad \determ{\Sigma}\,\tr{\Sigma^{-1}\rho}^2
  =(\cosh\vartheta-z\sinh\vartheta)^2\,.
\end{equation}
Then we have
\begin{eqnArray}\fl
  \label{eq:QB-3}
  (\D\rho)\,g(\rho)&=&\ds\D x\,\D y\,\D z \,\frac{1}{2\pi}
  \frac{n-1}{4^{n-1}}\frac{\Gamma(2n)}{\Gamma(n)^2}
  \frac{(1-x^2-y^2-z^2)^{n-2}}{(\cosh\vartheta-z\sinh\vartheta)^{2n}}
  \nonumber\\[2ex]\hspace*{-1em}&=&
  \ds\D s\,2(n-1)s(1-s^2)^{n-2}\;\frac{\D\phi}{2\pi}\,
  \;\frac{\D z}{2^{2n-1}}\,\frac{\Gamma(2n)}{\Gamma(n)^2}
     \frac{(1-z^2)^{n-1}}{(\cosh\vartheta-z\sinh\vartheta)^{2n}}\,,
\end{eqnArray}
where $s,\phi$ are polar coordinates in the unit disk and
${(x,y)=\sqrt{1-z^2}\,(s\cos\phi,s\sin\phi)}$.
In view of the factorization of the $s,\phi,z$ version, the probability
$(\D\rho)\,g(\rho)$ is the product of its three marginal probabilities, 
which tells us that $s$, $\phi$, and $z$  are independent random variables.%
\footnote{Note that the substitution
  $z=(z'+\tanh\vartheta)/(1+z'\tanh\vartheta)$ 
     yields a $z'$ marginal ${\propto\D z'\,(1-{z'}^2)^{n-1}}$
     that does not depend on $\vartheta$.
     Similarly, the substitution
     ${x=x'/\sqrt{(\cosh\vartheta)^2-(x'\sinh\vartheta)^2}}$ yields a
     $\vartheta$-independent $x'$
     marginal ${\propto\D x'\,(1-{x'}^2)^{n-1}}$, and likewise for $y$.}
This $g(\rho)$ is largest for $\peak$ with
$(x,y,z)_{\mathrm{peak}}^{\ }=(0,0,z_{\mathrm{peak}}^{\ })$ where
$z_{\mathrm{peak}}^{\ }$ is given by
\begin{equation}
  \label{eq:QB-4}
  \tanh\vartheta=\frac{(n-2)z_{\mathrm{peak}}^{\ }}{n-2z_{\mathrm{peak}}^2}\,,
\end{equation}
whereas the $z$-marginal is largest for the $z$ that solves
\begin{equation}
  \label{eq:QB-5}
  \tanh\vartheta=\frac{(n-1)z}{n-z^2}\,,
\end{equation}
which gives a value between $z=0$ and $z_{\mathrm{peak}}^{\ }$.

For the most natural choice of orthonormal basis matrices ---
that is ${B_1=2^{-\half}\sigma_x}$, ${B_2=2^{-\half}\sigma_y}$, and
${B_3=2^{-\half}\sigma_z}$ --- we have
\begin{equation}
  \label{eq:QB-6}
  \epsilon=\frac{1}{\sqrt{2}}
  \column[cc]{\varepsilon_3 &\varepsilon_1-\I\varepsilon_2\\
              \varepsilon_1+\I\varepsilon_2& -\varepsilon_3}
\end{equation}
here, and find
\begin{equation}
  \label{eq:QB-7}
  \log\frac{g(\peak+\epsilon)}{g(\peak)}
  \cong-\half\frac{4(n-2)}{1-z_{\mathrm{peak}}^2}
  {\left(\varepsilon_1^2+\varepsilon_2^2
      +\frac{1+\frac{2}{n}z_{\mathrm{peak}}^2}{1-z_{\mathrm{peak}}^2}
      \varepsilon_3^2\right)}
\end{equation}
near the peak.
The longitudinal variance (coordinate $\varepsilon_3$) is smaller by the
factor $\left(1-\peak[z]^2\right)/\left(1+\frac{2}{n}\peak[z]^2\right)$
than the transverse variance; 
all variances decrease when $n$ and $z_{\mathrm{peak}}$ increase.

\begin{figure}
  \centering
  \includegraphics{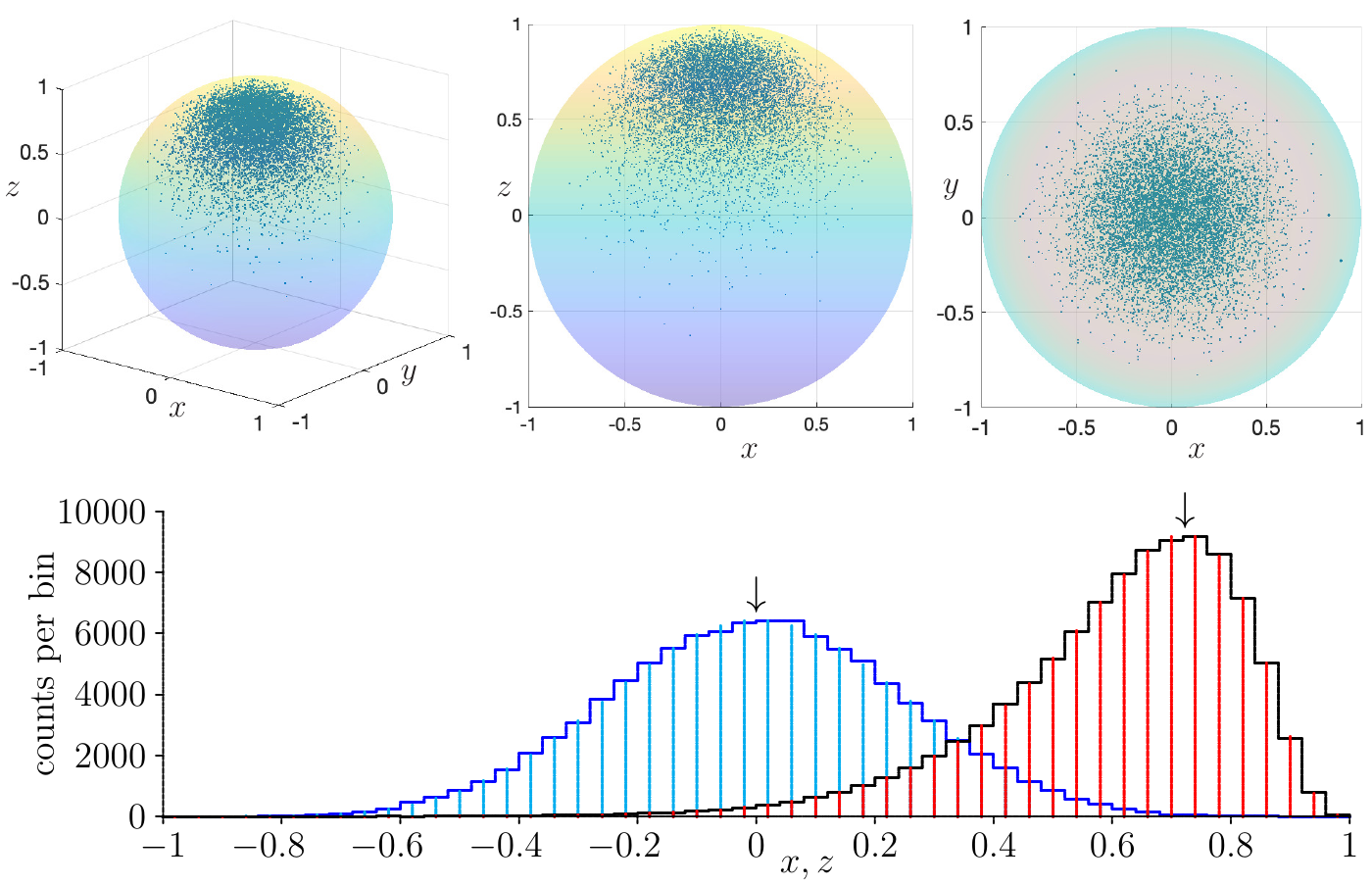}
  \caption{\label{fig:qubit} Single-qubit quantum Wishart distribution. 
    The top plots show a sample of $10\vts000$ quantum states drawn from
    $g(\rho)$ of \Eq{QB-3} for ${n=5}$ and
    ${\peak=\half(\unit_2+\frac{4}{5}\sigma_z)}$.
    Left: Pseudo-3D scatter plot of $g(\rho)$ inside the Bloch ball.
    Center: Scatter plot in the $x,z$ plane of $g(\rho)$ integrated over $y$.
    Right: Scatter plot in the $x,y$ plane of $g(\rho)$ integrated over $z$.
    \newline
    Bottom: Histograms of the $x$-marginal (blue staircase) and the
    $z$-marginal (black staircase), obtained by binning a sample of
    $100\vts000$ quantum states into fifty intervals of $0.04$.
    The impulses (cyan for $x$, red for $z$) show the expected values of counts
    in each bin.
    The arrows indicate the peak locations of the analytical marginals at
    ${x=0}$ and ${z=0.7223}$.}
\end{figure}

These matters are illustrated in figure~\ref{fig:qubit} for ${n=5}$ and
${\peak[z]=\frac{4}{5}}$, where we have scatter plots of $g(\rho)$ and of 
its $x,z$ and $x,y$ marginal distributions, of $10\vts000$ quantum states each.
There are also histograms, from a sample of $100\vts000$ quantum states,
for the one-dimensional marginal distributions in $x$ and $z$, which
we compare with the histograms computed from the analytical expressions.
The sample marginals are respectively obtained by ignoring the Bloch-ball
coordinate $y$, the coordinate $z$, the coordinate pair $(y,z)$, or the
coordinate pair $(x,y)$ of the sample entries.
This deliberate ignorance corresponds to integrating $g(\rho)$ over these
variables.
Note, in particular, the very good agreements of the sample histograms with
the analytical ones, which is partial confirmation that the sampling
algorithm of sections~\ref{sec:Wishart} and~\ref{sec:effSample} does indeed
yield a sample in accordance with the Wishart distribution in~\Eq{QB-3}.

\subsection{Example: Multi-qubit analog}\label{sec:multiQB}
As one multi-qubit example, we consider an analog of \Eq{QB-2}--\Eq{QB-4} for
$\nqb$ qubits,
\begin{eqnArray}
  \label{eq:nQB-1}
  \Sigma&\doteq&\ds\cosh\vartheta\;\unit_m
  +\sinh\vartheta\;\sigma_z^{\otimes\nqb}\,,
  \nonumber\\{}
  \peak&=&\ds\frac{1}{m}\Bigl(\unit_m\rule{0pt}{23pt}
      +z_{\mathrm{peak}}^{\ }\;\sigma_z^{\otimes\nqb}\Bigr)\,,
\end{eqnArray}
with ${m=2^{\nqb}}$ and $\vartheta$ related to $z_{\mathrm{peak}}^{\ }$
through the analog of \Eq{QB-4} with ``$2$'' replaced by $m$.
As observed in \Eq{shape-G}, different values of $z_{\mathrm{peak}}^{\ }$ do
not just give different peak locations, they also give different shapes of the
distribution. 
We look at $g(\rho)$ for
${\rho=\peak+\varepsilon m^{-\half}\sigma_z^{\otimes\nqb}}$, where
  $\varepsilon$ is the coordinate in the longitudinal direction. 
In other words, $\rho$ is of the form of $\peak$ in \Eq{nQB-1} with
$\varepsilon m^{\half}$ added to $z_{\mathrm{peak}}^{\ }$.

We quantify the width of the distribution, as a function of $\varepsilon$, by
the full width at half maximum (FWHM). 
The approximate value that follows from the gaussian approximation in
\Eq{peak-shape}, 
\begin{equation}
  \label{eq:peak-width}
  \mathalpha{\mathrm{FWHM}}
  \cong\frac{2}{m}{\left(1-z_{\mathrm{peak}}^2\right)}
  {\left[\frac{\log(4)}{(n-m){\left(1+\frac{m}{n}z_{\mathrm{peak}}^2\right)}}
    \right]}\power{\half}\,,
\end{equation}
catches the dependence on $z_{\mathrm{peak}}^{\ }$ and $n$ well.
It slightly overestimates FWHM for small $\peak[z]$ values, is very close to
the actual FWHM for ${\peak[z]\simeq0.5}$, and slightly underestimates FWHM
for large $\peak[z]$ values.
The relative error is in the range ${-4\%\,\cdots\, 2\%}$
when ${n>m+16/m}$; this is illustrated in figure~\ref{fig:FWHM-example}.

\begin{figure}
  \centering
  \includegraphics{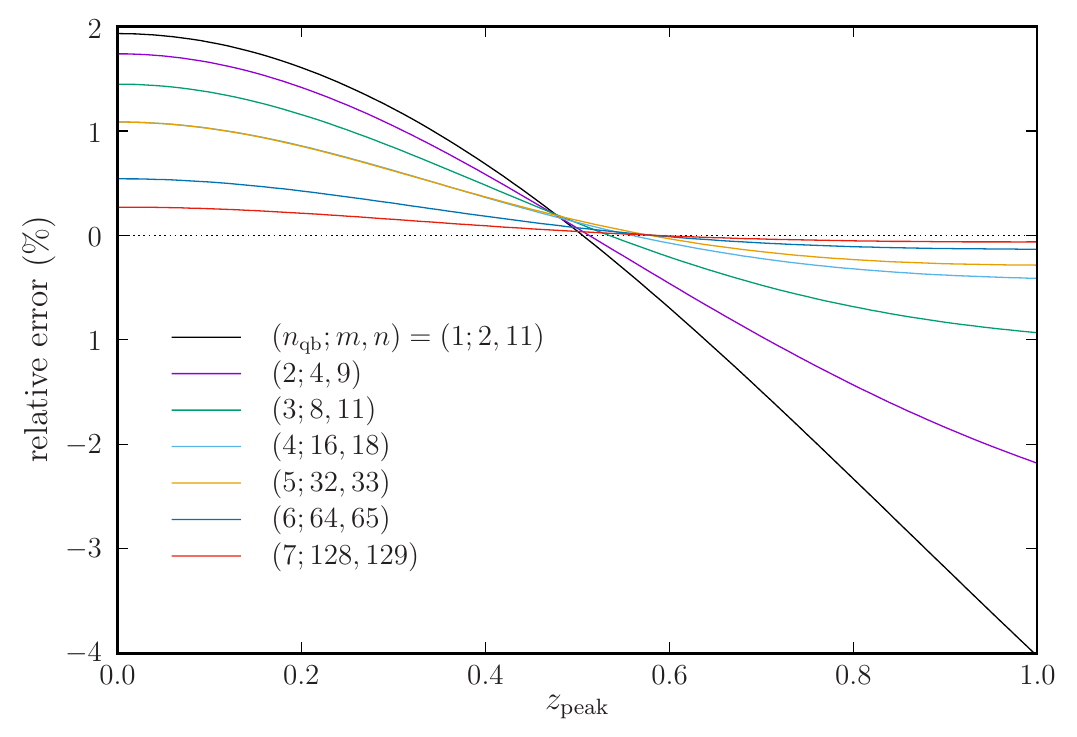}
  \caption{\label{fig:FWHM-example}Accuracy of the approximation
    \Eq{peak-width}.
    For one to seven qubits ($m=2^{\nqb}=2,4,8,\dots,128$), the plot reports
    the relative error for ${n=\lfloor m+16/m\rfloor+1}$ as a function of
    $\peak[z]$; the error is smaller for larger $n$ values.
    Note that the curves for ${\nqb=4}$ and ${\nqb=5}$ are indiscernible for
    ${\peak[z]\lesssim0.5}$.}
\end{figure}

\subsection{Linearly shifted distributions}\label{sec:linshift}
After choosing $\peak$ and thus $\Sigma$ such that the peak shape of the
Wishart distribution $g(\rho)$ fits to that of the target
distribution, in the sense discussed at the end of section~\ref{sec:target},  
the two peak shapes are matched.
The peak locations usually do not agree, however.

The proposal distribution is, therefore, obtained by shifting the Wishart
distri\-bution $g(\rho)$.
After acquiring a sample drawn from $\quW_m(n,\Sigma)$,
we turn each $\rho'$ from the sample into $\rho$ by the linear shift
map 
\begin{equation}
  \label{eq:shift}
  \rho=\rho'+\Delta\rho\quad\mbox{with}\quad (\D\rho)=(\D\rho')\,,
\end{equation}
where $\Delta\rho$ is a traceless hermitian $m\times m$ matrix that we choose
suitably. 
The distribution of the $\rho$ sample then peaks at ${\rho=\peak+\Delta\rho}$
and is characterized by the probability element 
\begin{equation}
  \label{eq:shift-g}
  (\D\rho)\,g_{\mathrm{s}}(\rho)
  =[\D\rho]\,\frac{\Gamma(mn)}{\Gamma_m(n)}
      \frac{\determ{\rho-\Delta\rho}^{n-m}}
        {\determ{\Sigma}^{n}\,\tr[10pt]{\Sigma^{-1}(\rho-\Delta\rho)}^{mn}}\,.
\end{equation}
Since the mapping \Eq{shift} linearly shifts the entire state space, 
there are $\rho\vts$s with no preimage $\rho'$ in the state space, so
that ${g_{\mathrm{s}}(\rho)=0}$ in the corresponding part of the state space.
There are also unphysical $\rho\vts$s with $g_{\mathrm{s}}(\rho)>0$; they will
be discarded during the rejection sampling discussed in
section~\ref{sec:accept-reject}.  
The matter is illustrated in figure~\ref{fig:linearPeakShift}.
While the fraction of physical $\rho\vts$s in the shifted distribution
$g_{\mathrm{s}}(\rho)$ is quite large ($\gtrsim0.6$) for a single qubit even
when the shift is by half of the radius of the Bloch ball, this fraction is
quite small for several-qubit distributions unless the shift itself is small
enough. 
Therefore, one needs to compromise between $\Sigma$, $n$, and $\Delta\rho$ and
exploit \Eq{optSigma} when choosing $\Sigma$ and $n$ such that $\Delta\rho$
is kept small.

\begin{figure}
  \centering
  \includegraphics{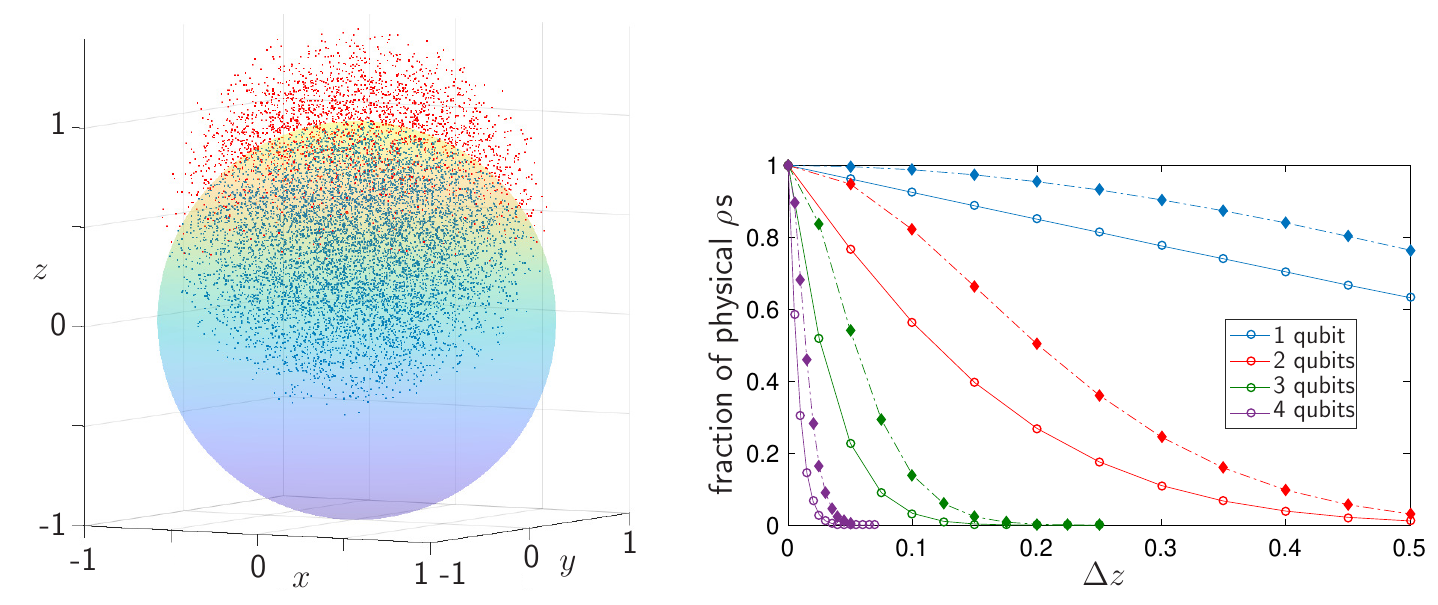}
\caption{\label{fig:linearPeakShift}%
  Linearly shifted Wishart distributions. 
  Left: Uniform single-qubit distribution shifted by
  ${\Delta\rho=\frac{1}{4}\sigma_z}$ and partly beyond the Bloch ball.
  The physical $\rho\vts$s in the sample of $10\vts000$ entries are
  marked by blue dots, the unphysical ones by red dots.
  Right: The fraction of physical $\rho\vts$s after shifting an isotropic
  $\nqb$-qubit distribution by
  ${\Delta\rho=\half\Delta z\,\sigma_z^{\otimes \nqb}}$ for $\nqb=1$, $2$,
  $3$, and $4$, for ${n=m=2^{\nqb}}$ (circles) and ${n=m+1}$ (diamonds).
  Since the distribution is narrower for larger $n$, the fraction of physical
  $\rho\vts$s is systematically larger for ${n=m+1}$ than for ${n=m}$.}
\end{figure}

The rejection sampling of section~\ref{sec:accept-reject} requires a proposal
distribution that is strictly positive and, therefore, we need to fill the
void created by the lack of $\rho'$s for certain $\rho\vts$s. 
For this purpose, we supplement the sample obtained by the shift \Eq{shift}
with $\rho\vts$s drawn from the uniform distribution $\quW_m(m,\unit_m)$; this
has no effect on the peak shape or the peak location. 
The full distribution of the proposal sample is then a convex sum of
$g_{\mathrm{s}}(\rho)$ and ${g(\rho)=\mathrm{constant}}$.
When ${0<\kappa<1}$ is the weight assigned to the uniformly distributed
subsample, and $N$ is the desired sample size, we repeat the sequence
\begin{equation}
  \label{eq:lottery}
  \parbox{280pt}{%
    \begin{tabular}{@{{\bf--}\ }p{160pt}@{}}
    draw $\Psi$ from $G(\Psi|\unit_m)\,;$\\[0.3ex]
      compute 
      $\ds\rho=\frac{\Sigma^{\half}\Psi\Psi\adj\Sigma^{\half}}%
      {\tr{\Sigma\Psi\Psi\adj}}+\Delta\rho\,;$
    \end{tabular}%
  }
\end{equation}
until we have $(1-\kappa)N$ entries in the sample.
Then we add $\kappa N$ $\rho\vts$s from the uniform distribution to complete
the proposal sample.

\enlargethispage{0.2\baselineskip} %%++

The proposal sample produced in this way is drawn from a distribution with the
probability element
\begin{eqnarray}\label{eq:proposal}
  (\D\rho)\,g_{\mathrm{s},\kappa}(\rho)
  &\propto&\ds[\D\rho]\,(1-\kappa)
      \frac{\Gamma(mn)}{\Gamma_m(n)}
      \frac{\determ{\rho-\Delta\rho}^{n-m}}
      {\determ{\Sigma}^{n}\,\tr[10pt]{\Sigma^{-1}(\rho-\Delta\rho)}^{mn}}
  \nonumber\\
  &&\ds\mbox{}+[\D\rho]\,\kappa\,\frac{\Gamma(m^2)}{\Gamma_m(m)}\,,
\end{eqnarray}
where we write ``$\propto$'' rather than ``$=$'' because we are missing the
overall factor that normalizes $g_{\mathrm{s},\kappa}(\rho)$ to unit integral
over the physical $\rho\vts$s.
The proposal distributions of this kind are nonzero for rank-deficient
$\rho\vts$s and the samples can have many $\rho\vts$s near a part of the
state-space boundary if $\peak+\Delta\rho$ is close to the boundary.
This is a useful feature when we need to mimic a target distribution that
peaks near or on the boundary, and the proposal sample
should have many points near the boundary where the unshifted $g(\rho)$
vanishes when ${n>m}$; see section~\ref{sec:examples} for examples.
Such target distributions assign very little probability to the void region of
the shift, and then we may not need many $\rho\vts$s from the
uniform distribution in the proposal sample.

The rejection sampling discussed in section~\ref{sec:accept-reject} has a high
yield only if the proposal distribution matches the target distribution well
and, therefore, a judicious choice of the parameters that define the sample
is crucial for the overall efficiency of the sampling algorithm.
In practice, we choose $n,\Sigma,\Delta\rho,\kappa$ and generate a small
sample, just large enough for a good estimate of the overall yield.
After trying out several choices, we generate the actual large sample for
the $n,\Sigma,\Delta\rho,\kappa$ with the largest yield.
Then we verify the sample by the procedure described in
section~\ref{sec:verify}.

\section{The target distributions}\label{sec:target}
The target distribution --- the probability distribution on the quantum state
space from which we want to sample --- can be of any form.
An important use of quantum samples is in Bayesian analysis, where
integrals over the high-dimensional quantum state space are central
quantities, which can often only be computed by Monte Carlo integration.
The samples are then drawn from the respective prior distribution,
or the posterior distribution after accounting for the knowledge provided by
the experimental data.
The target distributions that we use for illustration are of this kind.
The sampling algorithm can, of course, also be used for other target
distributions.

\subsection{Target distributions that arise in quantum state estimation}%
\label{sec:target1}%
To be specific, the examples used for illustration in
section~\ref{sec:examples} refer to the scenario of quantum state estimation
where many uncorrelated copies of the system are measured by an apparatus and
one of the $K$ outcomes is found for each copy.
The sequence of detection events constitute the data $D$ and the probability
of observing the actual data, if the system is prepared in the state $\rho$, is
the likelihood
\begin{equation}\label{eq:l-hood}
	L(D|\rho) = \prod_{k=1}^K p_k^{\nu_k}\,,
\end{equation}
where $p_k$ is the probability of observing the $k$th outcome and $\nu_k$
is the count of $k$th outcomes in the sequence; the counts themselves,
  ${\bnu=(\nu_1,\nu_2,\nu_3,\,\dots,\nu_K)}$, are a minimal statistic. 
The dependence on $\rho$ is given by the Born rule,
\begin{equation}\label{eq:Born}
	p_k=\tr{\rho\Pi_k}\,,
\end{equation}
with the probability operator $\Pi_k$ for the $k$th outcome.
The $\Pi_k$s are nonnegative and add up to the identity,
\begin{equation}
  \label{eq:POM}
  \Pi_k\geq0\,,\qquad\sum_{k=1}^K\Pi_k=1\,,
\end{equation}
but are not restricted otherwise.
The permissible probabilities $\vec{p}=(p_1,p_2,\dots,p_K)$ are those
consistent with ${\rho\geq0}$ and ${\tr{\rho}=1}$, they are a convex subset in
the probability simplex --- identified by $p_k\geq0$ and
${\sum_{k=1}^Kp_k=1}$ --- and it can be CPU-time expensive to check if a
certain $\vec{p}$ is permissible.

The target distribution $f(\rho)$ is the posterior distribution, the product
of the prior probability density $w_0(\rho)$ and the likelihood,
\begin{equation}\label{eq:target}
  f(\rho)\propto L(D|\rho)w_0(\rho)\,,
\end{equation}
where we do not write the proportionality constant that ensures proper
normalization.
Usually, it is expedient to use a conjugate prior, that is: $w_0(\rho)$ is of
the product form in \Eq{l-hood} with prechosen values for the $\nu_k$s, which
need not be integers, and then the posterior is also of this form.
For the purposes of this paper, therefore, we can just use a flat prior,
${w_0(\rho)=\textrm{constant}}$, for which the target distribution is the
likelihood in \Eq{l-hood}, possibly with noninteger values for the $\nu_k$s,
properly normalized to unit total probability when integrated over the $\rho$
space. 
With the volume element $(\D\rho)$ of \Eq{drho} and \Eq{drho'}, 
the probability element of the target distribution is then
\begin{equation}
  \label{eq:target-prob}
  (\D\rho)\,f(\rho)\propto (\D\rho)\,\prod_{k=1}^K\tr{\Pi_k\rho}^{\nu_k}
\end{equation}
for the physical values of the integration variables --- those for which
$\rho\geq0$ in \Eq{rho-varrho}, the convex set of quantum states ---
and $f(\rho)=0$ for all unphysical $\rho\vts$s.
Accordingly, the normalization integral
\begin{equation}
  \label{eq:target-norm}
  \int(\D\rho)\,f(\rho)=\INT{\rho}\,f(\rho)=1
\end{equation}
has no contribution from unphysical $\rho\vts$s.

The particular form of the target distribution, proportional to the likelihood
in \Eq{l-hood}, invites us to regard it as a function on the probability
simplex, a polynomial on the physical subset of permissible probabilities and
vanishing outside.
Indeed, the sampling algorithms in \cite{Shang2015,Seah2015} are random walks
in the probability simplex, and these algorithms have the practical
limitations mentioned above:
Either one needs to check if a candidate entry for the sample is permissible,
which has high CPU-time costs;
or one needs a complicated parameterization of the probability space and then
faces issues with the large Jacobian matrix, its determinant, and its
derivatives.
Therefore, alternative algorithms will be useful, such as schemes
that directly generate samples from the quantum state space rather than the
probability space associated with the~$\Pi_k$s.
It is the aim of this paper to contribute such an alternative algorithm.

We emphasize a crucial difference between the target distribution in
\Eq{target-prob} and the proposal distribution in \Eq{proposal}: While
both have analytical expressions, which will be important in section
\ref{sec:accept-reject}, the proposal distribution is defined by its sampling
algorithm, whereas we do not know an analogous sampling algorithm for the
target distribution.
Therefore, we cannot sample from the target distribution in a direct way and
must resort to processing samples drawn from the proposal distribution.

\subsection{Peak location}\label{sec:target2}
The target distribution $f(\rho)\propto L(D|\rho)$ is peaked at $\ML$, given by
\begin{equation}
  \label{eq:ML}
  \max_\rho\bigl\{L(D|\rho)\bigr\}=L(D|\ML)\,,
\end{equation}
which is to say that $\ML$ is the maximum-likelihood estimator for the data
$D$ \cite{Hradil1997,Paris2004}.
While it is possible that $L(D|\rho)$ is maximal for a multidimensional set of
$\ML$s, there is a unique $\ML$ for each of the target distributions that
we use for illustration.
If the relative frequencies ${\tilde\nu_k=\nu_k\Big/\sum_{k'=1}^K \nu_{k'}}$
are permissible probabilities, then ${\tilde\nu_k=\tr{\Pi_k\ML}}$
identifies $\ML$, otherwise one needs to determine $\ML$ numerically,
perhaps by the fast algorithm of \cite{Shang2017}, and one may find a
rank-deficient $\ML$ on the boundary of the state space.
When $\ML$ is on, or near to, the boundary it is usually more difficult to
sample from the state space in accordance with the target distribution.%
\footnote{One way of checking if certain probabilities $\vec{p}$ are
  permissible, is to regard them as relative frequencies of mock data and
  determine $\ML$ for these data.
  The probabilities are physical if ${p_k=\tr{\Pi_k\ML}}$ for all $k$, 
  and only then.}

Now, harking back to the final paragraph in section~\ref{sec:peak}, we note
that the option of matching a rank-deficient $\ML$ by the $\peak$ of a ${n=m}$
Wishart distribution does not work well usually, because $\quW_m(m,\Sigma)$
is then  maximal for all $\rho\vts$s in the range of $\ML$.
Therefore, we cannot get a good match when the rank of $\ML$ is larger than
one. 

When $\ML$ has full rank, the choice ${\Delta\rho=\ML-\peak}$ suggests itself
for the shift in section~\ref{sec:linshift}.
This is not a viable option, however, when $\ML$ is rank deficient,
a typical situation when the $\nu_k$s in \Eq{l-hood} are small numbers.
Rather, we choose $\Delta\rho$ such that $g_{\mathrm{s}}(\rho)$ peaks
for a full-rank $\rho$ that is close to $\ML$, as this gives a better over-all
yield; see the example in figure~\ref{fig:ar2qubitNC2_3} in
section~\ref{sec:examples}.

\subsection{Peak shape}\label{sec:target3}
Let us briefly consider the problem of maximizing the target
distribution $f(\rho)$ over all $\rho\vts$s of the form \Eq{rho-varrho}
\emph{without} enforcing ${\rho\geq0}$, which amounts to regarding $f(\rho)$ as
a function of the coordinates $\varrho_l$ and finding the maximum on the
coordinate space.
While the permissible $\rho\vts$s in \Eq{ML} are positive unit-trace
$m\times m$ matrices, we now maximize over all hermitian unit-trace matrices,
including those with negative eigenvalues.
This maximum is reached for ${\rho=\LINV}$.
Whenever $\LINV$ is in the quantum state space, ${\LINV\geq0}$, we have
${\LINV=\ML}$; and $\ML$ sits on the boundary of the state space whenever
$\LINV$ is outside the state space,
${\LINV\not\geq0}$.%
\footnote{The mock probabilities ${\widehat{p}_k=\tr{\Pi_k\LINV}}$ are always
  in the probability simplex; they are in the convex set of the physical
  probabilities only when ${\LINV=\ML}$.}

As noted above, we do not match the peak location of
$g_{\mathrm{s}}(\rho)$ with that of $f(\rho)$ when ${\ML\neq\LINV}$
is rank-deficient, whereas the shift ${\Delta\rho=\LINV-\peak}$
suggests itself when $\LINV$ is a quantum state.
It is then worth trying to match the peak shape of the quantum
Wishart distribution $g(\rho)$ with that of the target distribution $f(\rho)$. 
The expressions corresponding to \Eq{peak-shape} and \Eq{shape-G}, now for
$\rho\vts$s near ${\LINV=\ML}$ in the target distribution, are
\begin{equation}
  \label{eq:target-shape}
  \log\frac{f(\LINV+\epsilon)}{f(\LINV)}
  \cong-\half\sum_{k=1}^K\frac{\nu_k}{\tr{\Pi_k\LINV}^2}\tr{\Pi_k\epsilon}^2
  =-\half\sum_{l,l'}\varepsilon_l F_{ll'} \varepsilon_{l'}
\end{equation}
with
\begin{equation}
  \label{eq:shape-F}
  F_{ll'}=\sum_k\tr{B_l\Pi_k}\,\frac{\nu_k}
                               {\tr{\Pi_k\LINV}^2}\,\tr{\Pi_kB_{l'}}\,.
\end{equation}
When choosing the proposal distribution, one opts for $\peak$ and
$n$ such that the matrix $G$ resembles the matrix $F$.
It is usually not possible to get a perfect match because $G$ derives from a
Wishart distribution and is, therefore, subject to the symmetries discussed
after \Eq{g}, whereas $F$ is not constrained in this way.

In practice, we resort to looking at the one-dimensional ``longitudinal''
slice of $g(\rho)$ along the line from the completely mixed state
$m^{-1}\unit_m$ to $\peak$, that is ${\epsilon\propto\peak-m^{-1}\unit_m}$,
and choose the parameters such that this single-parameter distribution is a
bit wider%
\footnote{\label{fn:tails}%
  This illustrates the rule that one should ``sample from a
  density $g$ with thicker tails than $f$'' \cite{Wasserman2004}.}
than that of the corresponding longitudinal slice of the target
distribution $f(\rho)$, obtained for ${\epsilon\propto\LINV-m^{-1}\unit_m}$;
in example of section~\ref{sec:multiQB}, this slice is parameterized by the
coordinate increment $\varepsilon$.
The matter will be illustrated by more examples in section~\ref{sec:examples};
see figure~\ref{fig:ar2qubitNC2_3}.

Regarding the shift $\Delta\rho$ when ${\LINV\neq\ML}$, we note that
${\Delta\rho=r(\ML-\peak)}$ with ${r\lesssim1}$ is a good first try
for the trial-and-error search in the last paragraph of
section~\ref{sec:linshift}. 
In this situation, it is more important to match well the
distributions on and near the boundary of the state space than at $\ML$,
and we adjust $n$, $\Sigma$, $\kappa$, and $\Delta\rho$ in \Eq{lottery} for a
higher over-all yield.

\section{Rejection sampling}\label{sec:accept-reject}
We convert the proposal sample, drawn from the distribution
$g_{\mathrm{s},\kappa}(\rho)$, into a sample drawn from the target
distribution $f(\rho)$ by rejection sampling, a procedure introduced by
John von Neumann in 1951~\cite{vonNeumann51}.
It consists of a simple accept-reject step: A $\rho$ from the proposal sample
is entered into the target sample with a probability proportional to the ratio
$f(\rho)/g_{\mathrm{s},\kappa}(\rho)$.
Since we want to have the largest possible yield, while the acceptance
probability $P_{\mathrm{acc}}(\rho)$ cannot exceed unity, we choose
\begin{equation}\label{eq:accept-reject}
  P_{\mathrm{acc}}(\rho)=\frac{1}{C}\frac{f(\rho)}{g_{\mathrm{s},\kappa}(\rho)}
  \quad\mbox{with}\quad
  C=\max_{\rho}{\left\{\frac{f(\rho)}{g_{\mathrm{s},\kappa}(\rho)}\right\}}\,,
\end{equation}
where the maximum is evaluated over all $\rho\vts$s in the proposal sample.
The unphysical $\rho\vts$s in the proposal sample, obtained when the shift in
the second step of the state lottery \Eq{lottery} puts $\rho$ beyond the
boundary of the state space, are always rejected because ${f(\rho)=0}$ for
every unphysical $\rho$.

It is crucial here that we have the analytical expressions for $f(\rho)$ and
$g_{\mathrm{s},\kappa}(\rho)$ in \Eq{target-prob} and \Eq{proposal}, and it
does not matter that we often do not know the normalization factor needed in
\Eq{target-norm} or missing in \Eq{proposal};%
\footnote{Similarly, in the context of \Eq{accept-reject} we can put aside
  factors that the two terms in \Eq{proposal} have in common, such as the
  proportionality factor between $(\D\rho)$ and $[\D\rho]$ in \Eq{drho'}.}
it is permissible to replace $C$ by an upper bound on the maximal ratio, at
the price of a lower overall yield.
Note that, unless we have an independent way of computing an
upper bound on $C$,%
\footnote{Such cases require symmetries in the target distribution
      that match those of the proposal distribution.
      While it may be permissible to choose a Bayesian prior accordingly,
      the data-driven posterior will lack the symmetries.}
the rejection sampling cannot be done while we are entering
$\rho\vts$s into the proposal sample; we must first compose the whole proposal
sample, then determine the value of $C$, and finally perform the rejection
sampling. 

As remarked in section~\ref{sec:linshift}, the proper choice of the
parameters that specify the proposal distribution
$g_{\mathrm{s},\kappa}(\rho)$ is crucial for a good overall efficiency.
Here is one more aspect that one should keep in mind: If $\kappa$ is too
small, the ratio $f(\rho)/g_{\mathrm{s},\kappa}(\rho)$ will be largest where
both distributions are small --- in the tails of the target distribution and
where $g_{\mathrm{s}}(\rho)$ vanishes, in particular when $\rho$ has no
preimage $\rho'$ in \Eq{shift} or when the preimage is rank deficient.%
\footnote{The occurrence of rank-deficient preimages is an issue even when
  ${\Delta\rho=0}$ and there is no shift in \Eq{lottery}, but ${\kappa>0}$ is
  still required.}
When this happens, the rejection sampling has a low yield.
Therefore, we choose $\kappa$ large enough to avoid this situation.
This is an element in the overall-yield optimization mentioned in the final
paragraph of section~\ref{sec:linshift}.

\section{Sample verification}\label{sec:verify}
We exploit some tools, which were introduced in \cite{Shang13} for the
quantification of the errors in quantum state estimation, for a verification
of the target sample.
In this section, $f(\rho)$ refers to the $\rho$ dependence of the target
distribution in \Eq{target-prob} without assuming the proper normalization of
\Eq{target-norm}, and $\ML$ is the quantum state for which $f(\rho)$ is
maximal,
\begin{equation}
  \label{eq:f-ML}
  f(\ML)=\max_{\rho}\{f(\rho)\}\,.
\end{equation}
Then, the sets of quantum states defined by
\begin{equation}
  \label{eq:Rlambda}
  \mathcal{R}_{\lambda}={\left\{\rho|f(\rho)\geq\lambda f(\ML)\right\}}
\quad\mbox{with}\quad 0\leq\lambda\leq1
\end{equation}
are nested regions in the state space with
${\mathcal{R}_{\lambda}\subset\mathcal{R}_{\lambda'}}$ if ${\lambda>\lambda'}$;
$\mathcal{R}_{\lambda=1}$ contains only $\ML$ and
$\mathcal{R}_{\lambda=0}$ is the whole state space.
It is easy to check whether a particular $\rho$ is in $\mathcal{R}_{\lambda}$
or not.

For lack of a better alternative, we borrow the terminology from
\cite{Shang13} and call
\begin{equation}\label{eq:size-cred}
  \label{eq:size}
  s_{\lambda}^{\ }=\frac{\int_{\mathcal{R}_{\lambda}}(\D\rho)}
                       {\int_{\mathcal{R}_0}(\D\rho)}
  \quad\mbox{and}\quad
  c_{\lambda}^{\ }=\frac{\int_{\mathcal{R}_{\lambda}}(\D\rho)\,f(\rho)}
                       {\int_{\mathcal{R}_0}(\D\rho)\,f(\rho)}
\end{equation}
the size and the credibility of $\mathcal{R}_{\lambda}$, respectively,
and we evaluate both integrals by a Monte Carlo integration.%
\footnote{The term was coined by Stanis\l{}aw Ulam \cite{Metropolis1987}.
  The method dates back to 1949 \cite{Metropolis1949};
  for a comprehensive textbook exposition, see \cite{Evans2000},
  for example.}
By counting how many $\rho\vts$s from a uniform sample are in
$\mathcal{R}_{\lambda}$, we get a value for $s_{\lambda}^{\ }$ as the
fractional count;
likewise, by counting how many $\rho\vts$s from the target sample are in
$\mathcal{R}_{\lambda}$, we get a value for $c_{\lambda}^{\ }$.

The accuracy of the  $s_{\lambda}^{\ }$ and $c_{\lambda}^{\ }$ values is
determined by the sampling error and that is small if the sample is large
($10^6$ entries, say) and of good quality;%
\footnote{Sampling errors are discussed in section VI\,A\,2 in \cite{Gu2019},
  for example.}
see section~\ref{sec:curse}.
Since the uniform sample has no quality issues, it provides a reference for
the target sample through the identity
\begin{equation}
  \label{eq:link}
  c_{\lambda}^{\ }=\frac{\lambda s_{\lambda}^{\ }
                       +\int_{\lambda}^1\D\lambda'\,s_{\lambda'}^{\ }}
                       {\int_0^1\D\lambda'\,s_{\lambda'}^{\ }}\,,
\end{equation}
which links the credibility as a function of $\lambda$ to the size.
Note that the $s_\lambda$ values for the target distribution $f(\rho)$ as well
as the $c_\lambda$ values from \Eq{link} can be computed before any sampling
from the target distribution is performed.

Good agreement between the reliable credibility values provided
by \Eq{link} and those obtained by the Monte Carlo integration of
\Eq{size-cred} confirms that the target sample is of good quality.
More specifically, the quality assessment proceeds from regarding
$c_{\lambda}^{\ }$ from \Eq{link} as exact%
\footnote{\label{fn:bias}%
    When estimating $c^{\ }_{\lambda}$ from a large uniform
    sample, we are accepting a small negative bias that is, however, of no
    consequence here. The finite difference between the estimated and
    the actual values of $c^{\ }_{\lambda}$ do matter, however. See
    the Appendix for details.}
and the values obtained from the
target sample as estimates,
\begin{equation}
  \label{eq:cred-est}
  \estc=\frac{1}{N_{\mathrm{tgt}}}\sum_{k=1}^{N_{\mathrm{tgt}}}
  \Chi\BL f(\rho_k)>\lambda f(\ML)\BR\quad\textrm{with}
  \ \Chi(A)={\left\{\begin{array}{ll}
       1  & \mbox{if $A$ true}\\ 0  & \mbox{if $A$ false}
    \end{array}\right\}}\,,
\end{equation}
where the sum is over all $\rho_k$s in the target sample, which has
$N_{\mathrm{tgt}}$ entries.
This is an unbiased estimator, $\Expect{\estc}=c_{\lambda}^{\ }$,
with the variance 
$\ExpecT{\estc^2}-c^{2}_{\lambda}%
=c_{\lambda}^{\ }(1-c^{\ }_{\lambda})/N_{\mathrm{tgt}}$.  
For the comparison of $\estc$ %(from the target sample)
with $c_{\lambda}$,
% (inferred from $s_{\lambda}$, obtained from a huge uniform sample),
then, we use the mean squared error
\begin{equation}\label{eq:def-Q}
  Q=\int_0^1\D\lambda\,\BL\estc-c_{\lambda}\BR^2\,,
\end{equation}
which has the expected value
\begin{equation}\label{eq:expect-Q}
  \Expect{Q}=\int_0^1\D\lambda\;
  \ExpecT[1.4]{\BL\estc-c_{\lambda}\BR^2}
  =\frac{1}{N_{\mathrm{tgt}}}
  \int_0^1\D\lambda\,c_{\lambda}(1-c_{\lambda})
\end{equation}
and the variance
\begin{eqnArray}\fl\label{eq:var-Q}
 \rule{1em}{0pt}\ExpecT{Q^2}-\Expect{Q}^2&=&\ds
  \frac{2}{N_{\mathrm{tgt}}^2}\int_0^1\D\lambda\int_0^1\D\lambda'\,
   c_{<}^2(1-c_{>}^{\ })^2
   \\&&\ds\mbox{}+\frac{1}{N_{\mathrm{tgt}}^3}
   \int_0^1\D\lambda\int_0^1\D\lambda'\,
   c_{<}^{\ }(1-c_{>}^{\ })(1-4c_{<}^{\ }-2c_{>}^{\ }
   +6c_{<}^{\ }c_{>}^{\ })\,,
\end{eqnArray}
where $c_<^{\ }=\min\{c_{\lambda}^{\ },c_{\lambda'}^{\ }\}$
and  $c_>^{\ }=\max\{c_{\lambda}^{\ },c_{\lambda'}^{\ }\}$.
We consider the target sample to be of good quality if its $Q$ value
differs from  $\Expect{Q}$ by less than two standard deviations,
and of very good quality if the difference is less than one standard
deviation.\footnote{More sophisticated tests are possible, among them
    the Kolmogorov--Smirnov test \cite{Kolmogorov1933,Smirnov1948}; see
    section 15.4 in \cite{Wasserman2004}, for instance.
    We are not exploring such other possibilities here.} 
In practice, we calculate the $Q$ values with the approximate
$c_{\lambda}$ values obtained via \Eq{link} from the $s_{\lambda}$ values
that, in turn, we estimate from a large uniform sample, and the difference
between the approximate and the actual $c_{\lambda}$ adds an additional term to
$\Expect{Q}$ of \Eq{expect-Q}; see \Eq{A-17} in the appendix and figure
\ref{fig:CredSize1qubit2} in section~\ref{sec:examples}.

\section{The curse of dimensionality}\label{sec:curse}
As discussed in the preceding section, the fractional counts that give us
approximate values for $c^{\ }_{\lambda}$ are unbiased estimators for the
actual values with standard deviations of
$[c^{\ }_{\lambda}(1-c^{\ }_{\lambda})/N_{\mathrm{tgt}}]^{\half}$.
Likewise the approximate values for $s^{\ }_{\lambda}$ have standard
deviations of
$[s^{\ }_{\lambda}(1-s^{\ }_{\lambda})/N_{\mathrm{ufm}}]^{\half}$,
where $N_{\mathrm{ufm}}$ is the
total number of $\rho\vts$s in the uniform sample;
we take for granted that $N_{\mathrm{ufm}}$ is large enough that the sampling
error in $s^{\ }_{\lambda}$, and the propagated error in $c_{\lambda}^{\ }$ of
\Eq{link} can be ignored.
We now emphasize that the accuracy of these estimates is determined by the
sample size and is independent of the dimension of the quantum state space.
In this regard, then, these estimates do not fall prey to the ``curse of
dimensionality'' that Richard Bellman observed \cite{Bellman61}.
But we cannot really escape from the curse, which is known to affect all
major sampling methods,%
\footnote{All samples for higher-dimensional spaces that we are
      aware of, the samples in the repository \cite{QSampling} among them, are
      specified by the sampling algorithm, not by a target distribution.}
including the rejection sampling of section~\ref{sec:accept-reject},
and also occurs in other domains, such as optimization, function
approximation, numerical integration, and machine 
learning~\cite{Donoho2000}.

\enlargethispage{0.7\baselineskip} %%++

Let us first consider storage requirements.
We need 8 bytes of memory for one double precision number, which requires 24
bytes for one single-qubit state, 120 bytes for one two-qubit state, 504
bytes for one three-qubit state, 2040 bytes = 2.04\,kB for one four-qubit
state, and so forth, picking up a factor $\gtrsim4$ for each additional qubit:
growth proportional to $m^2$, the dimension of the state space.
This linear-in-$m^2$ increase in memory per quantum state is, however, paired
with an exponential decrease of the overall acceptance probability during the
rejection sampling.
Suppose we have a pretty good proposal distribution that is a 90\% match in
every dimension, then --- roughly speaking --- the overall match is
${0.9^3=0.729}$ for a single qubit, ${0.9^{15}=0.206}$ for a qubit pair,
${0.9^{63}=1.31\times10^{-3}}$ for three-qubit states, and a dismal
${0.9^{255}=2.15\times10^{-12}}$ for four-qubit states.
Accordingly, a target sample of $10^5$ states requires a proposal sample
stored in $3.3\,\mathrm{MB}$, $58\,\mathrm{MB}$, $38\,\mathrm{GB}$, and
$95\,\mathrm{EB}$, respectively.
Even if cleverly chosen proposal distributions can gain a few powers of $10$
on these rough estimates, it is clear that the curse of dimensionality
prevents us from generating useful target samples for many-qubit systems
by the rejection sampling of section~\ref{sec:accept-reject}.

%\enlargethispage{-0.5\baselineskip} %%++

Similarly, the CPU-cost increases with the dimension quite dramatically.
There is the positivity check that ensures ${f(\rho)=0}$ if ${\rho\not\geq0}$,
which has a CPU-cost proportional $m^3$ for a $m\times m$ matrix, and finding
the value of $C$ in \Eq{accept-reject} has a CPU-cost proportional to the size
of the proposal sample if we generously assume that the evaluation of
$g_{\mathrm{s},\kappa}(\rho)$ and $f(\rho)$ is CPU-cheap.
Clearly, the CPU-cost is forbiddingly large for many-qubit systems with
${m=2^{\nqb}}$.
We can, however, avoid the need for checking that ${\rho\geq0}$ by
choosing ${\Delta\rho=0}$ in \Eq{proposal}, while accepting the price of a
smaller yield of the rejection sampling or more steps of the evolution
algorithm.

The examples in section~\ref{sec:examples} demonstrate that, when using the
modified Wishart distribution of \Eq{proposal} as the proposal distribution,
we manage to sample efficiently \mbox{one-,} two-, and three-qubit states.
For four-qubit states, with our limited computational power, we can only
sample a certain class of simple target distributions by the rejection
sampling. 
Thus, although the sampling algorithm introduced here is superior to other
methods --- in particular, we obtain uncorrelated samples --- 
yet better methods for sampling from higher-dimensional state spaces
are in demand.
In a separate paper we combine Wishart distributions with sequentially
constrained Monte Carlo sampling \cite{DelMoral2006,Golchi2016} and so manage
to sample higher-dimensional quantum systems rather efficiently~\cite{Li2020}.

\section{Results}\label{sec:examples}
First, through examples of sampling qubit states, we will illustrate in detail
how to make use of the Wishart distribution for sampling quantum states with a
given target distribution.
Following that, we present examples of sampling %qutrit and
two-qubit states.
Although algorithms for reliably sampling such low-dimensional quantum states
exist (for example, the HMC sampling method works well for system with
dimension ${m<6}$~\cite{Shang2015, Seah2015}), our method has the advantage of
producing non-correlated samples.
Next, we apply our method to the sampling of three- and four-qubit systems
which are notoriously difficult for other sampling algorithms, and the method
introduced here, while applicable, suffers from a high rejection rate.

\subsection{Qubits}\label{sec:ex-1qb}
Assume that qubits are measured by a tetrahedral POM with the
probabilities\cite{Rehacek2004}
\begin{equation}\label{eq:tetraPOM}
  \begin{array}[b]{rclcrcl}
    p_1&=&\ds\frac{1}{4}\Bigl(1+\frac{x-y-z}{\sqrt{3}}\Bigr)\,,&\qquad&
    p_3&=&\ds\frac{1}{4}\Bigl(1+\frac{z-x-y}{\sqrt{3}}\Bigr)\,,\\[2ex]
    p_2&=&\ds\frac{1}{4}\Bigl(1+\frac{y-z-x}{\sqrt{3}}\Bigr)\,,&&
    p_4&=&\ds\frac{1}{4}\Bigl(1+\frac{x+y+z}{\sqrt{3}}\Bigr)\,,
  \end{array}
\end{equation}
where $\rho$ is parameterized as in \Eq{QB-1}.
Accordingly, we have ${K=4}$ in \Eq{l-hood}, \Eq{POM}, and \Eq{target-prob}.
For the observed counts of detection events $\bnu=(\nu_1,\nu_2,\nu_3,\nu_4)$,
the probability element of the target distribution is
\begin{equation}
  \label{eq:tetra1}
  (\D\rho)\,f(\rho)\propto\D x\,\D y\,\D z\,\eta\bigl(1-x^2-y^2-z^2\bigr)\,
  p_1^{\nu_1}p_2^{\nu_2}p_3^{\nu_3}p_4^{\nu_4}
\end{equation}
where the step function selects the permissible $x,y,z$ values, those of the
unit Bloch ball.
As noted above, we do not need to calculate the normalization factor, as it
plays no role in \Eq{accept-reject}.

Because the POM is symmetric --- technically speaking, it is a 2-design
\cite{Zauner2011} --- in the simplest, if untypical, scenario where
each outcome is observed equally often, the target
distribution is peaked at the completely mixed state
${\rho_{\mathrm{peak}}=\half\unit_2}$; $f(\rho)$ is not isotropic, however, as the
tetrahedron has privileged directions in the Bloch ball.  
We use the proposal distribution of \Eq{proposal} with ${m=2}$,
${\Sigma\doteq\unit_2}$, ${\Delta\rho=0}$, and ${\kappa>0}$, that is: an
isotropic Wishart distribution with an admixture of the uniform distribution;
the uniform-distribution component ensures ${g(\rho)>f(\rho)}$ in the tails.
The yield --- the acceptance rate of the rejection sampling --- depends on the
parameter $n$ that controls the width of the peak.
For example, a rather high acceptance rate of $P_{\mathrm{acc}}=60.8\%$ is
obtained with ${\kappa=0.1}$ and ${n=14}$ for ${\bnu=\{25,25,25,25\}}$;
by contrast, we have $P_{\mathrm{acc}}=33\%$ for ${n=18}$, which is thus a
worse choice for $n$ here.
The top and middle rows in figure~\ref{fig:fgOverlap1qubit} refer
to these $n$ values.

\begin{figure}[t]
\centerline{\includegraphics{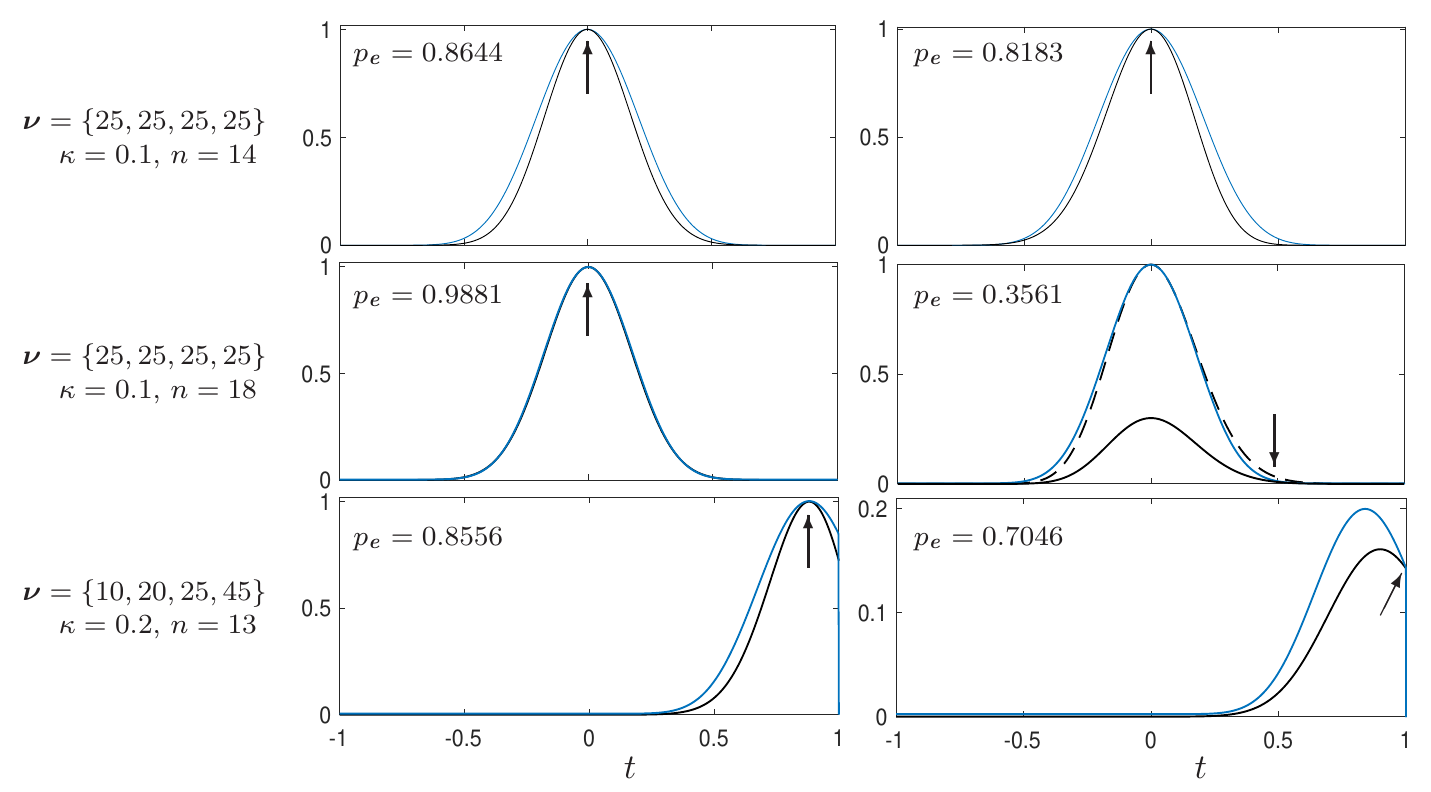}}
\caption{Line densities of the qubit target distribution
  $f(\rho)$ in \Eq{tetra1}  (black) and the proposal distribution
  $g_{\mathrm{s},\kappa}(\rho)$ in \Eq{proposal} with ${m=2}$ (blue)
  for ${\rho(t)=\half(\unit_2+t\vec{e}\cdot\bsigma)}$ with a unit vector
  $\vec{e}$.
  The target distributions (solid black) are rescaled by a factor of
  $\max_t\{f\bigl(\rho(t)\bigr)/g\bigl(\rho(t)\bigr)\}$ for each line, with
  the arrows pointing to where the blue and solid-black curves coincide;  
  the dashed black curve in the middle-right plot shows the target
  distribution rescaled for matched peaks.
  In each plot, the value of $p_{\vec{e}}$ is the acceptance rate of
  \Eq{t-accrate}, the ratio of the areas under the solid black and blue
  curves.
  \textbf{Top row and middle row:}
  The target distribution is for ${\bnu=\{25,25,25,25\}}$, and the isotropic
  proposal distributions have ${\Sigma=\unit_2}$ and ${\Delta\rho=0}$ as well as
  $\kappa=0.1$ and $n=14$ or $18$, respectively. 
  The unit vectors $\vec{e}$ are chosen at random.
  The overall acceptance rates are $P_{\mathrm{acc}}=60.8\%$ (top row) and
  $P_{\mathrm{acc}}=33\%$ (middle row).
  \textbf{Bottom row:}
  The target distribution is for $\bnu=\{10,20,25,45\}$, the proposal
  distribution has $\Sigma=\unit_2$, $\kappa=0.2$, $n=13$, and
  $\Delta\rho\neq0$
  such that the peaks of the target and proposal distributions are at the same
  location; the overall acceptance rate is $P_{\mathrm{acc}}=28.6\%$.
  In the left plot, the unit vector $\vec{e}$ points from the center of the
  Bloch ball to the peak location, whereas we have a randomly   chosen
  $\vec{e}$ in the right plot; note the smaller range of values in the right
  plot.} 
  \label{fig:fgOverlap1qubit}
\end{figure}

In figure~\ref{fig:fgOverlap1qubit} we exploit our knowledge of the exact
forms of the target distribution $f(\rho)$ and the proposal distribution
$g(\rho)$ for a comparison of the distributions along diameters of
the Bloch ball.
More specifically, the qubit states considered are
$\rho(t)=\half(\unit_2+t\vec{e}\cdot\bsigma)$ with a unit vector $\vec{e}$ and
${-1\leq t\leq1}$.
Accordingly,
the acceptance rate for this one-parameter family of states is
\begin{equation}
  \label{eq:t-accrate}
  p_{\vec{e}}=  \max_t\Biggl\{\frac{f\Lp\rho(t)\Rp}
                                  {g\Lp\rho(t)\Rp}\Biggr\}^{-1}
              \frac{\mathop{\scalebox{1.3}{$\int$}}_{\!\!\! -1}^1\D t\,
                   f\Lp\rho(t)\Rp}
                 {\mathop{\scalebox{1.3}{$\int$}}_{\!\!\! -1}^1\D t\,
                   g\Lp\rho(t)\Rp}\,.
\end{equation}
We note in passing that suitable analogous expressions work also for higher
dimensional systems.

As the top and middle rows in figure~\ref{fig:fgOverlap1qubit} confirm, we
have a better overall yield for $n=14$ than for $n=18$;
for $n=18$, where the width of the Wishart distribution well matches that of
the target distribution, the almost
perfect yields for some directions $\vec{e}$, as in the middle-left example,
does not compensate for the low yield for other diameters, as in the
middle-right example.
The middle-right plot also indicates why $n=18$ is a poor choice,
as the maximum of $f\Lp\rho(t)\Rp/g\Lp\rho(t)\Rp$ occurs in the tails of the
distributions, not near the peak of $f\Lp\rho(t)\Rp$; recall footnote
\ref{fn:tails} in this context.

In the typical experimental scenario, the counts of measurement clicks are
not balanced among the POM outcomes, the target distribution does not peak at
the center of the Bloch ball, and is not approximately isotropic.
We use ${\bnu=\{10,20,25,45\}}$ to illustrate this situation in the bottom
row of figure~\ref{fig:fgOverlap1qubit}; here the peak of $f(\rho)$
is at distance $0.8832$ from the center of the Bloch sphere.
We can match the peak of the Wishart distribution by either changing the
covariance matrix $\Sigma$ in accordance with~\Eq{optSigma}, or by a suitable
shift $\Delta\rho$, or by a combination of both.
We find that ${\Sigma=\unit_2}$ in conjunction with ${\Delta\rho=\peak}$
is very effective as this provides a high acceptance rate.
For the event counts stated above and this shift, an acceptance rate of
${P_{\mathrm{acc}}=28.6\%}$ is achieved by using a Wishart sample with
${\kappa=0.2}$ and ${n=13}$ --- this is, in fact, a rather high
acceptance rate in view of the many $\rho$s in the Wishart sample that the
shift renders unphysical.  
We use a larger admixture of the uniform sample here as we need to fill in the
void left behind after the shift by $\Delta\rho$.
The bottom row in figure~\ref{fig:fgOverlap1qubit} shows the distribution
along the diameter through the peak location (left) and along another,
randomly chosen, diameter (right).

\begin{figure}[p]
\centerline{\includegraphics[viewport=-5 0 410 451,clip=]%
    {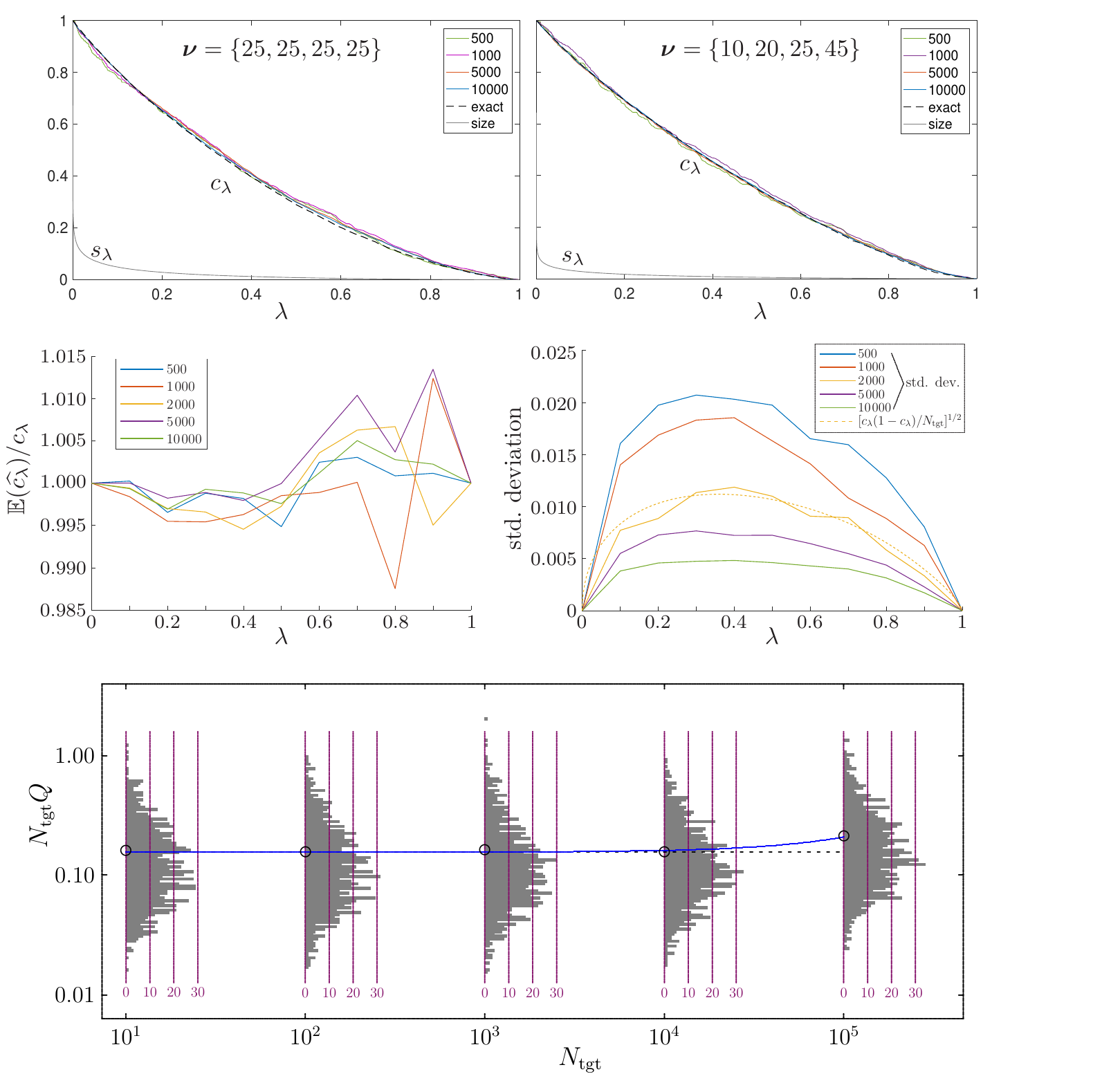}}
\caption{\label{fig:CredSize1qubit2}%
  Verification for qubit samples.
  \textbf{Top row:}
  Samples generated for data ${\bnu=\{25,25,25,25\}}$
  (left) and ${\bnu=\{10,20,25,45\}}$ (right).
  The size $s_\lambda^{\ }$ and the regarded-as-exact credibility
  $c_{\lambda}^{\ }$ of \Eq{link} are evaluated with the aid of uniformly
  distributed samples of (${N_{\mathrm{ufm}}=10^7}$).
  The estimated credibility $\estc$ is obtained from target
  samples with $N_{\mathrm{tgt}}=500$, $1\vts000$, $5\vts000$, or $10\vts000$
  entries.
  \textbf{Middle row:} 
  Ratio of the expected and the exact value of $\estc$ (left, for a single
  target sample);
  standard deviation of $\estc$ (right, averaged over $100$ target samples);
  for $\bnu=\{10,20,25,45\}$ and target samples with
  ${N_{\mathrm{tgt}}=500}$, $1\vts000$, $2\vts000$, $5\vts000$,
  or $10\vts000$ entries each.
  There are values for $\lambda=0.0, 0.1, 0.2, \dots,1.0$, connected by
  straight lines that guide the eye. The smooth dashed yellow curve on the
  right shows the square root of the variance
  for ${N_{\mathrm{tgt}}=2\vts000}$; see after \Eq{cred-est}.
  \textbf{Bottom:}
  Histograms of $N_{\mathrm{tgt}}Q$ values for $1\vts000$ samples of
  $N_{\mathrm{tgt}}=10^1$, $10^2$, $10^3$, $10^4$, and $10^5$ for
  ${\bnu=\{10,20,25,45\}}$. 
  The purple vertical lines indicate $0$, $10$, $20$, or $30$ counts in the
  histogram bins. The circles show the mean values for each histogram, to
  which the blue curve is fitted in accordance with \Eq{A-17}, with $0.15567$
  (dashed horizontal line) and $5.245\times10^{-7}$ for the values of the two
  integrals. 
}
\end{figure}

To verify the sample we evaluate the size and credibility of the bounded
likelihood regions as described in section~\ref{sec:verify}.
The size $s_{\lambda}$ is estimated from a uniform sample with ten million
entries.
The regarded-as-exact $c_{\lambda}$ values then obtained from \Eq{link} are
used for reference (dashed black curves in figure~\ref{fig:CredSize1qubit2}, top
row).
We numerically evaluate $c_\lambda$ using \Eq{cred-est} for
$N_{\mathrm{tgt}}=500$, $1\vts000$, $5\vts000$, and $10\vts000$
and compare it with the regarded-as-exact values in
the top row in figure~\ref{fig:CredSize1qubit2}.
The middle row in figure~\ref{fig:CredSize1qubit2}, with data for target sample
sizes of $N_{\mathrm{tgt}}=500$, $1\vts000$, $2\vts000$, $5\vts000$, and
$10\vts000$, shows that our method of sampling is reliable and the standard
deviation of $c_{\lambda}$ is proportional to $N_{\mathrm{tgt}}^{-\half}$ as
expected.
The bottom plot in figure~\ref{fig:CredSize1qubit2} confirms \Eq{A-17} in the
appendix, which corrects \Eq{expect-Q} by accounting for the difference
between the regarded-as-exact $c_{\lambda}$ values and the actual ones.
We note further that the standard deviations of the histograms are
approximately equal to their mean values (indicated by circles), so that the
criterion stated after \Eq{var-Q} tells us that all samples with $Q$ values
[cf.\ \Eq{def-Q}] less than twice the mean value are of very good quality.

\subsection{Qubit pairs}\label{sec:ex-2qb}
For simulated data for measurements on qubit pairs, we use the probabilities 
of a double tetrahedral measurement with the sixteen probability operators
\begin{equation}\label{eq:tetra-tetra}
\Pi_{k}=\Pi_l^{(1)}\otimes\Pi_m^{(2)}\quad\mbox{with\ } l,m=1,2,3,4\,,
\end{equation}
where $\Pi_l^{(1)}$ and $\Pi_m^{(2)}$ make up the single-qubit tetrahedron
POMs for the probabilities in \Eq{tetraPOM}, and the
index $k=4(l-1)+m$ covers the integers in the rage ${1\leq k\leq16=K}$.
For measurement data $\bnu=\{\nu_1,\nu_2,\dots,\nu_{16}\}$, the target
distribution has the form of \Eq{target-prob},
$(\D\rho)\,f(\rho)\propto(\D\rho)\,\prod_{k=1}^{16} \tr{\Pi_k\rho}^{\nu_k}$.

\enlargethispage{-0.7\baselineskip} %%++

First, we compare the performance of a proposal sample from the uniform
distribution with one from the Wishart distribution for the centered, symmetric
target distributions that refer to fictitious measurements with the same
number of events for each outcome, that is
$\nu_1=\nu_2=\cdots=\nu_{16}=\overline{\nu}$; see figure
\ref{fig:ar2qubitCentred}.
The acceptance rate $P_{\mathrm{acc}}$ obtained from using a uniformly
distributed proposal sample decreases exponentially as the measurement count
$\overline{\nu}$ increases and the target distribution becomes correspondingly
narrower.
By contrast, when we admix a sample drawn from the Wishart distribution, the
optimal acceptance rate decreases to about $0.1\%$ at around
$\overline{\nu}=20$ and then it increases slowly to about $1\%$ for
$\overline{\nu}=100$ and remains at around this value for even
larger~$\overline{\nu}$s. 
For example, an acceptance rate of $P_{\mathrm{acc}}=0.48\%$ is obtained by
adding $20\%$ of $\quW_4(6,\unit_4)$ for $\overline{\nu}=10$,
$P_{\mathrm{acc}}=0.091\%$ is obtained by adding $50\%$ of
$\quW_4(8,\unit_4)$ for $\overline{\nu}=20$,
and $P_{\mathrm{acc}}=0.64\%$ is obtained by adding $90\%$ of
$\quW_4(35,\unit_4)$ for $\overline{\nu}=100$.
Whereas, the acceptance rates for the uniform proposal samples are
$P_{\mathrm{acc}}=5.4\times10^{-4}$, $2.3\times10^{-5}$, and
$1.0\times10^{-8}$ for $\overline{\nu}=10$, $20$ and $100$, respectively.
The use of a proposal sample from the Wishart distribution clearly gives a much
larger acceptance rate for such a peaked target distribution than the uniform
proposal sample.

\begin{figure}[t]
\centerline{\includegraphics{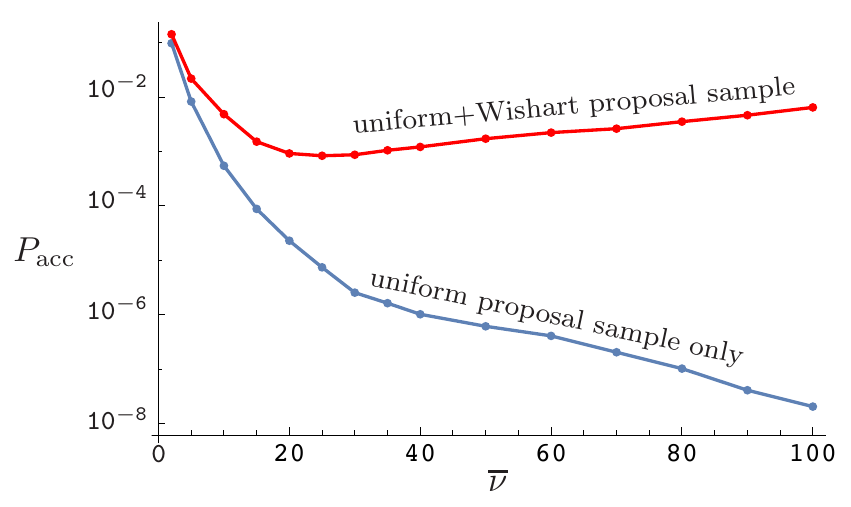}}
\caption{\label{fig:ar2qubitCentred}%
    Acceptance rate for sampling two-qubit states with target distribution for
    the data $\bnu=\{\nu_k=\overline{\nu};k=1,2,\dots,16\}$.
    The blue curve and the red curve show the acceptance as a function of
    $\overline{\nu}$ with a uniformly distributed proposal sample and the
    optimal result given by adding in a suitable Wishart distribution,
    respectively.
    Each data point of the acceptance rate is obtained with $10^8$ proposal
    samples.}
\end{figure}

While the use of Wishart distribution can increase the acceptance rate, to
reach the optimal acceptance rate one needs to find the most suitable Wishart
distribution to use.
Fortunately, this optimization is not difficult because one observes that the
optimal number of columns $n_{\mathrm{opt}}$ scales linearly with the total
number of counts; see figure~\ref{fig:ar2qubitCentredNcol}.
Moreover, we find that this proportionality of $n_{\mathrm{opt}}$ and $N$
holds quite generally when one samples non-centered target distributions.
When $N$ is large, the acceptance rate can be rather robust against the choice
of $n$ because the peak width depends only weakly on this parameter;
see section~\ref{sec:shape}. 

\begin{figure}[t]
\centerline{\includegraphics{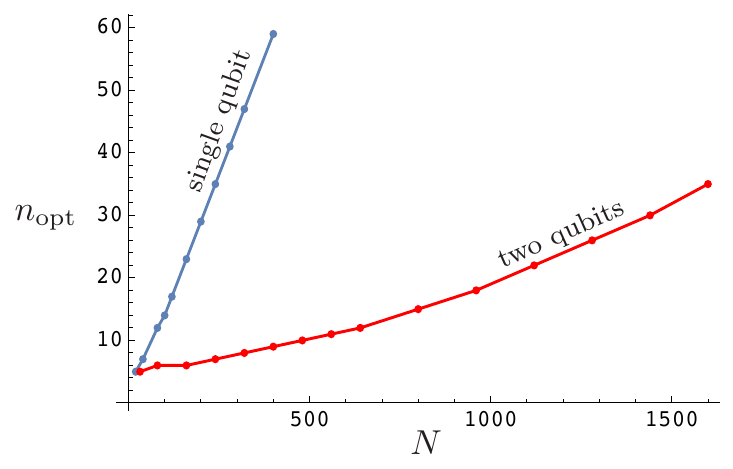}}
\caption{\label{fig:ar2qubitCentredNcol}%
    The optimal number of columns $n_{\mathrm{opt}}$ of the Wishart
    distribution for sampling one qubit (blue) and two qubit (red) centered
    target distributions against the total number of measurements $N$.}
\end{figure}

Next, let's check the performance of the method for sampling non-centered
target distributions.
For illustration, we use a target distribution given by a particular random
data obtained from $100$ measurements with
$\bnu=\{10, 4, 6, 4, 7, 6, 5, 6, 5, 6, 10, 6, 5, 6, 8, 6\}$.
The peak of the target distribution in the probability simplex corresponds to
a non-physical state, and the peak in the physical space is given by the
maximum-likelihood estimator $\ML$ which is a rank-3 state with eigenvalues
$\{0.5033,0.3377,0.1589,0\}$. 

%\newpage  %%++

As discussed in section~\ref{sec:linshift}, we can adjust the peak location of
the proposal distribution by a suitable shift.
We generate the proposal distribution by the two steps in~\Eq{lottery}.
First, we chose a state $\peak'=x_1\ML+\frac{1}{4}(1-x_1)\unit_4$ to be the
peak of the Wishart distribution $\quW_4(n,\Sigma)$ and find the corresponding
$\Sigma$ to produce a preliminary proposal sample.
Then, we shift the sample states by $\Delta\rho=x_2(\ML-\frac{1}{4}\unit_4)$
as in \Eq{shift}, after which we have a sample that is peaked at
$\peak=(x_1+x_2)\ML+\frac{1}{4}(1-x_1-x_2)\unit_4$; to this we admix a
fraction $\kappa$ from the uniform sample and arrive at a proposal sample in
accordance with the distribution $g_{\mathrm{s},\kappa}(\rho)$ of \Eq{proposal}.
When $x_1+x_2=1$, the peak of the proposal distribution coincides with the
maximum-likelihood estimator.
However, we do not need the peaks to coincide exactly, instead, they just need
to be close enough to give a good acceptance rate.

\begin{figure}[t]
\centerline{\includegraphics{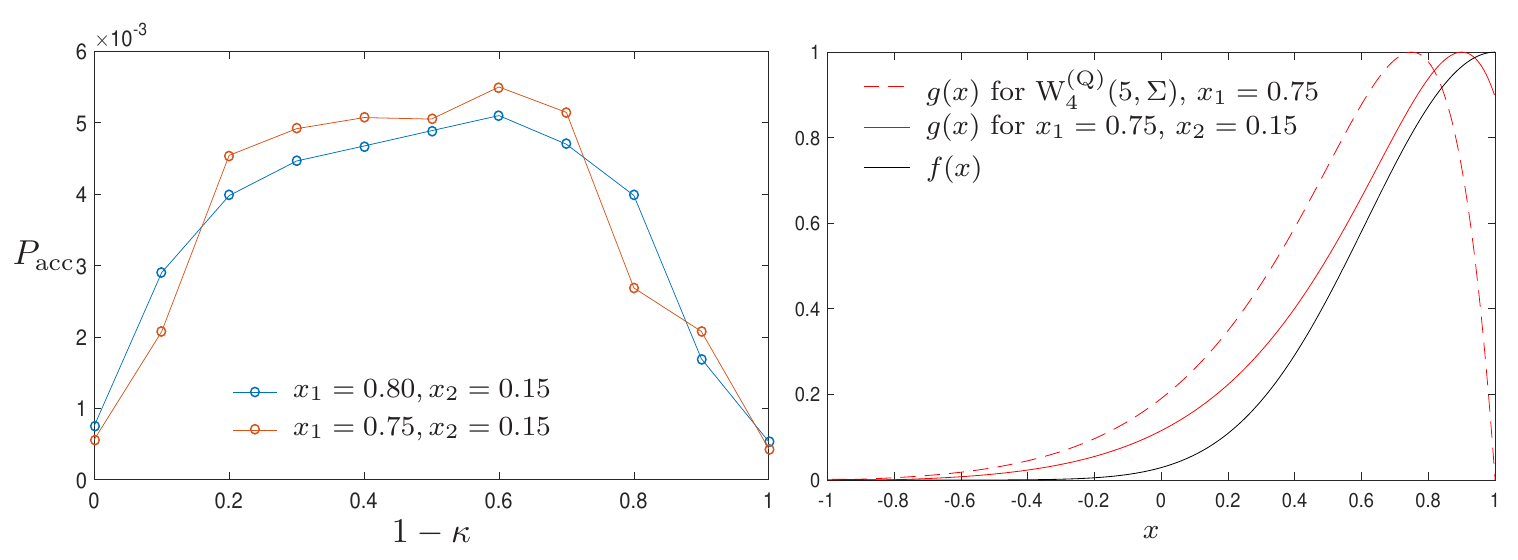}}
\caption{\label{fig:ar2qubitNC2_3}%
    \textbf{Left:} The acceptance rate $P_{\mathrm{acc}}$ against $1-\kappa$
    for a proposal sample with $10^8$ states obtained from a shifted
    non-centered Wishart distribution $\quW_4(5,\Sigma)$.
    \textbf{Right:} The cross-section density of the reference and target
    distributions for states of the form of
    $\rho(x)=x\ML+\frac{1}{4}(1-x)\unit_4$. }
\end{figure}

\begin{figure}[b]
\centerline{\includegraphics{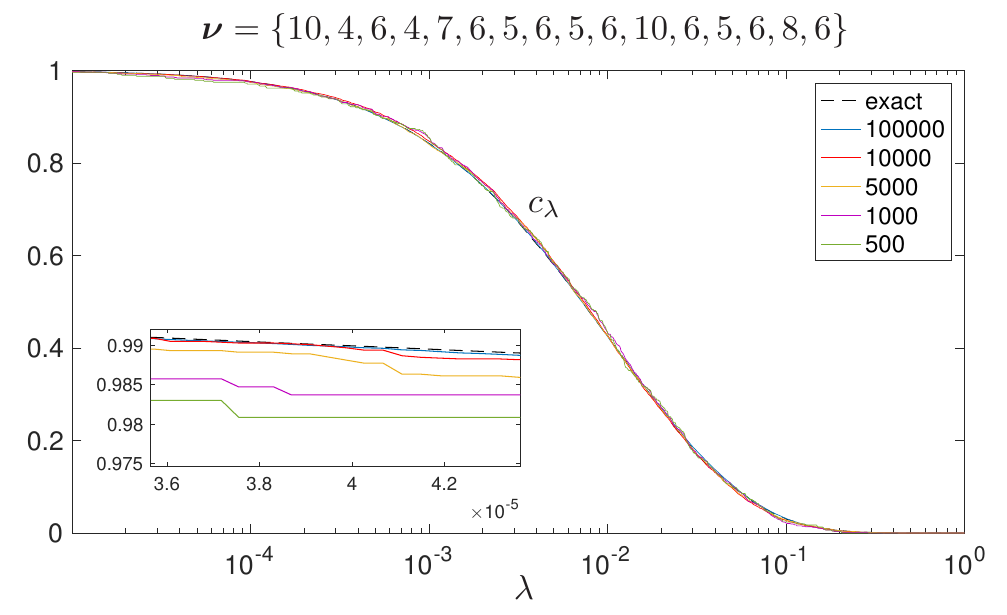}}
\caption{\label{fig:Cred2qubit2}%
    The credibility $c_\lambda$ evaluated for different sample sizes. $10^6$
    sample states from the uniform distribution are used to evaluate
    the size and the `exact' values of the credibility given by the dashed
    curve.} 
\end{figure}

For example, we achieve an  acceptance rate of ${P_\mathrm{acc}>0.5\%}$ with a
proposal distribution obtained for ${\kappa=0.6}$, ${n=5}$, ${x_1=0.75}$, 
and ${x_2=0.15}$; see figure~\ref{fig:ar2qubitNC2_3}.
For the verification of the sample, we evaluate the credibility $c_\lambda$
for different sample sizes and compare it to the regarded-as-exact value
computed from $s_\lambda$; see figure~\ref{fig:Cred2qubit2}.
The result shows that our method is reliable for sampling two-qubit systems
and $10\vts000$ sample points are enough for the estimation of physical
quantities, such as the credibility, with an error less than 0.1\%.
For this particular example, the acceptance rate for a uniformly
distributed proposal distribution is about 0.06\% which is smaller by about a
factor of ten.
Thus, using the Wishart distribution does improve the acceptance rate but the
improvement is not significant enough when applied to a target distribution
that is not narrowly peaked, such as the one for ${N=100}$ measurements.
The gain in efficiency is much more remarkable when the target distributed is
more peaked.
For instance, if the target distribution is given by ${N=1\vts000}$ measurements
with the same frequency of clicks for each POM outcome, an acceptance rate of
larger than $10^{-5}$ is achieved with a proposal distribution drawn
from $\quW_4(10,\Sigma)$ with ${x_1=0.8}$ and ${x_2=0.15}$.
This is at least $10^3$ times more efficient than using a uniform proposal
distribution which fails to produce any accepted sample point for as many as
$10^8$ proposal states.

\begin{table}[t]
  \caption{\label{tbl:CPU}%
  CPU time taken for generating a sample of $10^5$ states with different
  sampling strategies.
  All the tasks are CPU-parallelized and run on the same regular desktop.
  (a) Two-qubit sample with ${\nu_k=10}$ for ${k=1,\dots,16}$;
  (b) two-qubit sample with ${\bnu=\{10, 4, 6, 4, 7, 6, 5, 6, 5,
    6, 10, 6, 5, 6, 8, 6\}}$;
  (c) two-qubit sample with each $\nu_k$ equal to ten times the value of
  example (b); 
  (d) three-qubit sample with ${\nu_k=10}$ for ${k=1,\dots,64}$;
  (e) and (f) three- and four-qubit samples with a randomly generated count
  sequence of ${3\vts000}$ experiments.\footnotemark
  \newline
  Remarks:~\parbox[t]{320pt}{$\star\;$the acceptance rate is too low, less
    than $1$ in $10^8$,\\
  -- the strategy is not reliable,\\
  $^\dag\;$the acceptance rate is about $1.2\times10^{-5}$,\\
  $^{\dag\dag}\;$the acceptance rate is $21$ samples from
  $1.5\times10^8$ proposal samples.}}
\flushright\small
\begin{tabular}{@{\;}lrrcccc@{\;}}\br
  &\multicolumn{3}{c}{\underline{\rule{35pt}{0pt}two qubits\rule{35pt}{0pt}}}
  &\multicolumn{2}{c}{\underline{\rule{18pt}{0pt}three qubits\rule{16pt}{0pt}}}
  & \underline{\ four\vphantom{q}\ }\\[1ex]
sampling strategies & \multicolumn{1}{c}{(a)} &\multicolumn{1}{c}{(b)}
    &(c)&(d)&(e)&(f) \\ \mr
uniform [$\kappa=1$ in \Eq{proposal}]
                   & 60\,min  & 30\,min & $\star$
                   & $\star$& $\star$ & $\star$ \\ 
Wishart-uniform [$0<\kappa<1$]
                   &7\,min & 5\,min & 31\,h &100\,h$\vts^\dag$
         & 13\vts000\,h$\vts^{\dag\dag}$ & $\star$ \\[0.5ex] 
HMC~\cite{Shang13} & 80\,min & 80\,min &  -- & -- & -- & --\\
SCMC+Wishart~\cite{Li2020} &2.5\,min & 2.5\,min & 2.5\,min
      & 9\,min & 60\,min & 60\,h \\ \br
\end{tabular}
%{\small Text: AA bb cc.}
\end{table}
\footnotetext{\newcommand{\Z}{\phantom{0}}\label{fn:counts3qubits}%
The simulated data for the three-qubit example are
\begin{displaymath}\begin{array}{r@{}l}
  \bnu=\{&36,13,64,71,\ 14,16,\Z7,15,\ 60,10,\Z84,63,\ 64,\Z9,55,71,\\
         &\Z8,12,10,16,\ 16,48,67,62,\ \Z9,64,\Z75,63,\ 10,74,60,73,\\
         & 65,14,62,66,\ \Z9,57,76,53,\ 82,78,128,22,\ 61,44,25,27,\\
         & 56,12,52,66,\ 14,76,56,78,\ 45,47,\Z22,27,\ 66,68,25,102\}
\end{array}\end{displaymath}
with ${N=3\vts000}$.
The $64$ counts refer to the three-qubit analog of \Eq{tetra-tetra}.
For these data, $\ML$ is rank deficient.}

When sampling from the two-qubit state space, the CPU time taken when using
the Wishart distribution for the proposal is significantly lower than that for
other methods. 
Table~\ref{tbl:CPU} shows that for sampling a centered distribution with
${N=160}$ and a non-centered distribution with ${N=100}$ it is a few times more
efficient to use the Wishart distribution with an admixture of the uniform
distribution than solely the uniform distribution;
see the top-two entries in columns (a) and (b).
The advantage of using the Wishart distribution can become much more prominent
for sampling more peaked distributions with larger~$N$; see column (c).
Our method is also more efficient than the Hamiltonian Monte Carlo (HMC)
method that is discussed in \cite{Shang13}; see the third row. 
While the efficiency of the HMC method does not depend on $N$,
which can give it an advantage over the sampling from a uniform distributions
for large $N$, the current implementation of the HMC algorithm
is only reliable for systems with low dimension,
owing to issues with the stability of the algorithm; note also that
HMC yields correlated samples even if the correlations are usually weaker than
those in samples from other Markov chain MC methods.
In the fourth row in table~\ref{tbl:CPU} we list the CPU time for generating
samples with the sequentially constrained MC (SCMC) algorithm that we describe
in \cite{Li2020}; it appears to outperform all other methods.
If we consider the one-qubit and two-qubit situations of columns
(a) and (b), however, the ``Wishart-uniform method'' of this paper is
just as practical and requires much less effort in writing and debugging
computer code; also, the ``SCMC+Wishart method'' of reference \cite{Li2020}
builds on the foundations laid by the ``Wishart-uniform method.''

\subsection{Three and four qubits}\label{sec:ex-3+4qb}
The sampling method introduced here works for systems of any dimension --- at
least, there are no reasons of principle why it shouldn't.
In practice, however, the `curse of dimensionality' can become a serious
obstacle, due to limitations in both CPU time and memory; see
section~\ref{sec:curse}. 

The storage aspects discussed in section~\ref{sec:curse} is much more a
concern for three-qubit systems than for one-qubit or two-qubit systems.
When sampling three-qubit states, we use up to a maximum number of
$2.4\times10^8$ proposal states in each run of the algorithm, which takes up
about 120\,GB of storage.
The first example we investigated is a centered distribution for
$\bnu=\{\nu_k=10 \ \mathrm{for}\ k=1,\dots,64\}$.
No useful sample can be produced by drawing from a uniform distribution as its
acceptance rate is smaller than $10^{-8}/2.4$.
On the other hand, when sampling with a proposal distribution composed 20\% of
the Wishart distribution $\quW_{8}(9,\unit_8)$ and 80\% of the uniform
distribution --- ${\kappa=0.8}$ in \Eq{proposal} --- the acceptance rate is
approximately $1.2\times10^{-5}$. 
This allows us to obtain a sample size of $10^5$ within 100~hours.

Reliable samples for non-centered and/or narrower distributions can also be
generated from the Wishart-plus-uniform proposal distribution \Eq{proposal}.
For example, for the target distribution that corresponds to the
simulated data in footnote~\ref{fn:counts3qubits}, we obtain
${P_\mathrm{acc}\approx1.4\times10^{-7}}$ using 40\% of
$\quW_{8}(80,\unit_8)$, with shift parameters ${x_1=0.6}$ and ${x_2=0.35}$,
plus 60\% of the uniform distribution. 
It takes about 2.75~hours to produce 21 sample states from $1.5\times10^8$
proposal states.
Thus, to generate $10^5$ samples it would take about 13\vts000 hours,
as entered in column (e) of table~\ref{tbl:CPU}.  

For a four-qubit system, storage is even more of an issue; we can only store
about $6\times10^7$ states in 120 GB of memory.
We tried to sample a target distribution for ${N=3\vts000}$ simulated counts
from a Wishart distribution but failed to produce a reliable sample.
This suggests that the acceptance rate is well below $1.6\times10^{-8}$ and,
therefore, to sample states of such high dimension efficiently, yet other methods
are to be sought.
One option is the adaptation of the SCMC sampler \cite{DelMoral2006,Golchi2016}
to the sampling of quantum states.
We managed to produce uncorrelated samples of high-dimensional quantum states
in this way.
The ``SCMC+Wishart algorithm'' is, however, beyond the scope of this paper; we
deal with it in \cite{Li2020}.

\section{Summary and outlook}\label{sec:summary}
We established the probability distribution of the unit-trace, positive square
matrices that represent quantum states, generated from the gaussian
distribution for nonsquare matrices and the induced intermediate Wishart
distribution for positive matrices.
These distributions of quantum-state matrices are shifted suitably and
supplemented with an admixture of the uniform distribution so that they can
serve as tailored proposal distributions for the target distribution from
which we want to sample.

The target sample is generated from a proposal sample by an
accept-reject step.
Since the proposal sample is uncorrelated (or i.i.d.), so is the target
sample. 
This is a notable advantage over random walk algorithms, including that of
Hamiltonian Monte Carlo, which generate correlated samples with a
correspondingly reduced effective sample size.

The sampling algorithm works very well for samples from the one-qubit and
two-qubit state spaces, whereas the efficiency in generating three-qubit
samples is low, owing to the curse of dimensionality.
We could not generate a useful sample of four-qubit states drawn from a
structured target distribution. (Drawing from the uniform distribution or
other highly symmetric distributions is efficient.)

The discussion here is limited to the Wishart distributions that derive from
zero-mean gaussian distributions.
We explore the options offered by gaussian distributions with a nonzero mean
in a companion paper \cite{Bagchi2020} but the increase in flexibility does not
overcome the curse of dimensionality.

The general strategy explored in this paper --- equip yourself with a large,
CPU-cheap, uncorrelated proposal sample and turn it into a smaller target
sample by rejection sampling --- does not require a proposal distribution of the
Wishart-uniform kind for its implementation.
Other proposal distributions are possible and could very well yield
substantially larger acceptance rates.
While we have no particular suggestions to make, we trust that others will
till this field.

There is, fortunately, solid evidence that the sequentially constrained Monte
Carlo sampling algorithm succeeds in overcoming the curse of dimensionality to
some extent.
Rather than accepting or rejecting the quantum states in the proposal sample,
the sample as a whole is processed and turned into a proper target sample step
by step; this method also benefits from using a tailored Wishart distributions
for the proposal.
We present this approach in a separate paper \cite{Li2020}.

\section*{Acknowledgments}
S.B. is supported by a PBC postdoctoral fellowship at Tel-Aviv University.
The Centre for Quantum Technologies is a
Research Centre of Excellence funded by the Ministry of
Education and the National Research Foundation of Singapore.

\appendix
\section*{Appendix}\setcounter{section}{1}
In this appendix, we elaborate on footnote~\ref{fn:bias}.
As the analog of \Eq{cred-est}, we have
\begin{equation}\label{eq:A-1}
  \ests=\frac{1}{N_{\mathrm{ufm}}}\sum_{k=1}^{N_{\mathrm{ufm}}}
  \Chi\BL f(\rho_k)>\lambda f(\ML)\BR
  =\frac{1}{N_{\mathrm{ufm}}}\sum_{k=1}^{N_{\mathrm{ufm}}}
  \Chi\bigl(\lambda <\lambda_k\bigr)
\end{equation}
when estimating $s^{\ }_{\lambda}$ from the $N_{\mathrm{ufm}}$ entries
$\rho_k$ in the uniform sample, with
\begin{equation}\label{eq:A-2}
  \lambda_k=\frac{f(\rho_k)}{f(\ML)}\,,\quad 0\leq\lambda_k\leq1\,.
\end{equation}
The $\lambda_k$s are independent (uncorrelated) random variables, each
with the probability element 
\begin{equation}\label{eq:A-3}
  \Expect{\lambda'<\lambda_k<\lambda'+\D\lambda'}
  =\D\lambda'\,\biggl(-\pdi{\lambda'}{s_{\lambda'}}\biggr)\,,
\end{equation}
so that, for each $\lambda_k$,
\begin{equation}\label{eq:A-4}
  \ExpecT{\Chi(\lambda<\lambda_k)}
  =\int_{\lambda}^1\D\lambda'\,\biggl(-\pdi{\lambda'}{s_{\lambda'}}\biggr)
  =s_{\lambda}^{\ }\,.
\end{equation}
It follows that $\ests$ is unbiased,
$\Expect{\ests}=s_{\lambda}^{\ }$, and its variance is proportional to
$N_{\mathrm{ufm}}^{-1}$,
\begin{equation}
  \label{eq:A-5}
  \ExpecT{\ests^2}-\Expect{\ests}^2=\frac{1}{N_{\mathrm{ufm}}}
  s_{\lambda}^{\ }\BL1-s_{\lambda}^{\ }\BR\,.
\end{equation}
Further, the numerator in \Eq{link} is
\begin{eqnArray}\fl\rule{3em}{0pt}
  \label{eq:A-6}\ds
  \lambda s_{\lambda}^{\ }+\int_{\lambda}^1\D\lambda'\, s_{\lambda'}^{\ }
  &=&\ds\int_{\lambda}^1\D\lambda'\,
   \biggl(-\pdi{\lambda'}{s_{\lambda'}}\biggr)\lambda'
   =\int_0^1\D\lambda'\,
   \biggl(-\pdi{\lambda'}{s_{\lambda'}}\biggr)\Chi(\lambda<\lambda')
   \lambda'\\[2ex]
  &=&\ds\ExpecT{\Chi(\lambda<\lambda_k)\lambda_k}\,,
\end{eqnArray}
and
\begin{equation}
  \label{eq:A-7}
  \lambda \ests
  +\int_{\lambda}^1\D\lambda'\, \widehat{s_{\lambda'}}
  =\frac{1}{N_{\mathrm{ufm}}}\sum_{k=1}^{N_{\mathrm{ufm}}}
    \Chi\bigl(\lambda <\lambda_k\bigr)\lambda_k\,,
\end{equation}
is the corresponding unbiased estimator for the numerator; 
for ${\lambda=0}$, this is the estimator for the denominator,
\begin{equation}
  \label{eq:A-8}
  \int_0^1\D\lambda'\, \widehat{s_{\lambda'}}
  =\frac{1}{N_{\mathrm{ufm}}}\sum_{k=1}^{N_{\mathrm{ufm}}}\lambda_k\,.
\end{equation}Their ratio,
\begin{equation}%\fl%\rule{3em}{0pt}
  \label{eq:A-9}
  \estc^{(\mathrm{ufm})}=
  \frac{\lambda \ests
    +\sint_{\!\!\lambda}^1\D\lambda'\, \widehat{s_{\lambda'}}}
    {\sint_{\!\!0}^1\D\lambda'\, \widehat{s_{\lambda'}}}=
  \frac{\ssum\limits_{k=1}^{N_{\mathrm{ufm}}}
    \Chi\bigl(\lambda <\lambda_k\bigr)\lambda_k}
  {\ssum\limits_{l=1}^{N_{\mathrm{ufm}}}\lambda_l}\,,
  %=\sum_k\int\limits_0^{\infty}\D\alpha\,
  %\Chi\bigl(\lambda <\lambda_k\bigr)\lambda_k
  %\,\Exp{-\alpha\ssum_l\lambda_l}\,,
\end{equation}
is what we get from \Eq{link} for the regarded-as-exact credibility.

As we shall now demonstrate, $\estc^{(\mathrm{ufm})}$ has a negative
bias.
First, we note that
\begin{eqnArray}\label{eq:A-10}
  \ExpecT{\estc^{(\mathrm{ufm})}}
  &=&\ds\Expect{\sum_k
    \Chi\bigl(\lambda <\lambda_k\bigr)\lambda_k
    \int\limits_0^{\infty}\D\alpha\,\Exp{-\alpha\ssum_l\lambda_l}}\\
  &=&\ds\sum_k\int\limits_0^{\infty}\D\alpha\,\Expect{
    \Chi\bigl(\lambda <\lambda_k\bigr)\lambda_k\,\Exp{-\alpha\lambda_k}}
    \,\prod_{l(\neq k)}\Expect{\Exp{-\alpha\lambda_l}}\\
    &=&\ds N_{\mathrm{ufm}}\int\limits_0^{\infty}\D\alpha\,
    \Expect{\Chi\bigl(\lambda <\lambda'\bigr)\lambda'\,
      \Exp{-\alpha\lambda'}}
    \,\Expect{\Exp{-\alpha\lambda'}}^{N_{\mathrm{ufm}}-1}\,,
\end{eqnArray}
where $\lambda'$ is a random variable with the probability element of \Eq{A-3}
and the expected values refer to this random variable.
Then, we integrate by parts and arrive at
\begin{equation}\label{eq:A-11}
    \ExpecT{\estc^{(\mathrm{ufm})}}= c_{\lambda}^{\ }
  +\int\limits_0^{\infty}\D\alpha\,
  \Expect{\Exp{-\alpha\lambda'}}^{N_{\mathrm{ufm}}}
  \pdi{\alpha}{} C_{\lambda}(\alpha)\,,
\end{equation}
where 
\begin{equation}
  \label{eq:A-12}
  C_{\lambda}(\alpha)=\frac{\ExpecT{\Chi(\lambda<\lambda')\lambda'
                      \,\Exp{-\alpha\lambda'}}}
                  {\Expect{\lambda'\,\Exp{-\alpha\lambda'}}}
                  =\left(1+\frac{\ExpecT{\Chi(\lambda>\lambda')\lambda'
                          \,\Exp{-\alpha\lambda'}}}
                      {\ExpecT{\Chi(\lambda<\lambda')\lambda'
                          \,\Exp{-\alpha\lambda'}}}\right)^{\!-1}
\end{equation}
is a monotonic decreasing function of $\alpha$, and $C_{\lambda}(0)%
=\Expect{\lambda'}^{-1}\ExpecT{\Chi(\lambda<\lambda')\lambda'}%
=c^{\ }_{\lambda}$ is taken into account.
Accordingly, the term added to $c^{\ }_{\lambda}$ in \Eq{A-11} is negative,
that is: $\estc^{(\mathrm{ufm})}$ has a negative bias.
Indeed, when $\alpha'>\alpha$, we observe that
\begin{eqnArray}
  \label{eq:A-13}
  C_{\lambda}(\alpha')^{-1}&=&\ds1+
\frac{\ExpecT{\Chi(\lambda>\lambda')\lambda'
  \,\Exp{-\alpha\lambda'}\,\Exp{(\alpha'-\alpha)(\lambda-\lambda')}}}
   {\ExpecT{\Chi(\lambda<\lambda')\lambda'
   \,\Exp{-\alpha\lambda'}\,\Exp{-(\alpha'-\alpha)(\lambda'-\lambda)}}}
\\[3ex]&>&\ds1+\frac{\ExpecT{\Chi(\lambda>\lambda')\lambda'
                          \,\Exp{-\alpha\lambda'}}}
                      {\ExpecT{\Chi(\lambda<\lambda')\lambda'
                      \,\Exp{-\alpha\lambda'}}}
  =C_{\lambda}(\alpha)^{-1}\,,
\end{eqnArray}
or $C_{\lambda}(\alpha')<C_{\lambda}(\alpha)$.

Repeated integrations by parts express the bias in \Eq{A-11} as a sum of terms
proportional to powers of $N_{\mathrm{ufm}}^{-1}$.
The leading term is exhibited here:
\begin{equation}\fl\rule{1em}{0pt}
  \label{eq:A-14}\int\limits_0^{\infty}\D\alpha\,
  \Expect{\Exp{-\alpha\lambda'}}^{N_{\mathrm{ufm}}}
  \pdi{\alpha}{} C_{\lambda}(\alpha)
              =\frac{1}{N_{\mathrm{ufm}}}
              \frac{\ExpecT{{\lambda'}^2}}{\Expect{\lambda'}^2}
              {\left[c_{\lambda}^{\ }
                  -\frac{\ExpecT{\Chi(\lambda<\lambda'){\lambda'}^2}}
                      {\ExpecT{{\lambda'}^2}}\right]}+\cdots
\end{equation}
with
\begin{equation}\fl\rule{4em}{0pt}
  \label{eq:A-15}
  \ExpecT{\Chi(\lambda<\lambda'){\lambda'}^2}
  =\lambda^2 s_{\lambda}^{\ }
  +2\int_{\lambda}^1\D\lambda'\,\lambda'\,s_{\lambda'}^{\ }\,,\quad
  \ExpecT{{\lambda'}^2}=\ds
  2\int_0^1\D\lambda'\,\lambda'\,s_{\lambda'}^{\ }\,.
\end{equation}

Since both the variance of $\ests$ and the bias in
$\estc^{(\mathrm{ufm})}$ are proportional to $N_{\mathrm{ufm}}^{-1}$, 
the bias in $\estc^{(\mathrm{ufm})}$ is smaller than the standard deviation of
$\ests$, from which  $\estc^{(\mathrm{ufm})}$ is computed, by a factor of
$N_{\mathrm{ufm}}^{-\half}$.
Accordingly, this bias is of no consequence for the considerations in
section~\ref{sec:verify}, the more so if we recall the typical circumstances
of $N_{\mathrm{ufm}}\gg N_{\mathrm{tgt}}\gg1$.

In practice, we are using the estimate $\estc^{(\mathrm{ufm})}$ of \Eq{A-9},
obtained from a large uniform sample, in the place of the actual $c_{\lambda}$
when computing $Q$ of \Eq{def-Q}, that is
\begin{equation}
  \label{eq:A-16}
  Q\to\int_0^1\D\lambda\,{\left[\BL\estc-c_{\lambda}\BR
      -\BL\estc^{(\mathrm{ufm})}-c_{\lambda}\BR\right]}^2\,;
\end{equation}
the expected value of ${\estc^{(\mathrm{ufm})}-c_{\lambda}}$ (with respect to
the uniform distribution) is the bias in \Eq{A-11}.
As a consequence, the expected value of \Eq{expect-Q} (with respect to the
target distribution) acquires a small
additional term that does not depend on the size of the target sample,
\begin{equation}
  \label{eq:A-17}
  \Expect{Q}\to
  \frac{1}{N_{\mathrm{tgt}}}
  \int_0^1\D\lambda\,c_{\lambda}(1-c_{\lambda})
  +\int_0^1\D\lambda\,\BL\estc^{(\mathrm{ufm})}-c_{\lambda}\BR^2\,,
\end{equation}
and the variance in \Eq{var-Q} also acquires a corresponding additional term.

\end{document}